\documentclass[10pt,aps,prc,twocolumn,showpacs,superscriptaddress,floatfix,nofootinbib]{revtex4-1}

\usepackage{graphicx}  
\usepackage{dcolumn}   
\usepackage{bm}        
\usepackage{amssymb}   
\usepackage{amsfonts}
\usepackage{graphics}
\usepackage{epsfig}
\usepackage[normalem]{ulem}
\usepackage[usenames]{color}
\usepackage{multirow}
\usepackage{color}  
\usepackage{lipsum}
\usepackage{lineno}
\usepackage[noend]{algorithmic}
\usepackage[noend]{algorithm2e}

\def\pt{\mbox{$p_{\rm T}$}}
\def\kT{\mbox{$k_{\rm T}$}}
\def\KTThree{\mbox{$K_{\rm T,3}$}}
\def\KTFour{\mbox{$K_{\rm T,4}$}}
\def\qinv{\mbox{$q$}}
\def\qo{\mbox{$q_{\rm out}$}}
\def\qs{\mbox{$q_{\rm side}$}}
\def\ql{\mbox{$q_{\rm long}$}}

\DeclareMathSymbol{\mlq}{\mathord}{operators}{``}
\DeclareMathSymbol{\mrq}{\mathord}{operators}{`'}

\makeatletter

\newcommand{\Rmnum}[1]{\expandafter\@slowromancap\romannumeral #1@}
\makeatother

\begin{document}          

\title{Techniques for multiboson interferometry}
\author{Dhevan Gangadharan}
\email[]{dhevan.raja.gangadharan@cern.ch}
\affiliation{Nuclear Science Division, Lawrence Berkeley National Laboratory, 1 Cyclotron Rd, Berkeley, CA 97420}

\begin{abstract}
The quantum statistics (QS) correlations of identical bosons are well known to be sensitive to the space-time extent and dynamics of the particle emitting source in high-energy collisions.  
While two-pion correlations are most often experimentally measured, the QS correlations of three-pions and higher are rarely explored.  
A set of techniques to isolate and analyze 3- and 4-pion QS correlations is presented.  
In particular, the technique of built correlation functions allows one to more easily study the effects of quantum coherence at finite relative momenta instead of at the unmeasured intercept of correlation functions.

\end{abstract}
\maketitle

\section{Introduction}

In high-energy collisions, the quantum statistics (QS) correlations of identical particles provide the most sensitive probe of the space-time structure of the particle-emitting source \cite{Goldhaber:1960sf,Kopylov:1975rp,Gyulassy:1979yi}.
Most often, 2-pion Bose-Einstein correlations are experimentally measured for which the techniques are well established.
However, the statistical precision of modern experiments has opened the door to measurements of higher order QS correlations.

Higher order QS correlations contain additional information about the source which cannot be learned from 2-particle correlations alone.
In particular, the suppression expected from quantum coherence of identical pions in the final state increases substantially for higher orders.
The possibility of quantum coherence in high energy collisions has been considered several times before \cite{Andreev:1992pu,Plumer:1992au,Ornik:1993gb,Bjorken:1993cz,Bjorken:1997re,Akkelin:2001nd}.

One such mechanism is the disoriented chiral condensate (DCC) \cite{Bjorken:1993cz,Bjorken:1997re,Rajagopal:1997au}.
The chiral condensate can be characterized with four scalar fields ($\sigma,\vec{\pi}$) and is non-zero at low temperatures where chiral symmetry is spontaneously broken.
In the core of the collision, the chiral condensate is expected to vanish and the symmetry is restored.
A hot expanding shell of particles essentially shields the core from the true vacuum outside the shell.
As the core energy density drops, the chiral condensate forms and may be temporarily disoriented in one of the $\vec{\pi}$ directions instead of the true vacuum $\sigma$ direction.
The disoriented vacuum eventually reorients in the $\sigma$ direction and is accompanied by coherent pion radiation.  
The DCC is very different from a Bose-Einstein condensate for which the latter develops from critically large pion densities.

A quantum coherent or pure state of matter can be a rather delicate state of matter.  
Before the collision of two heavy-ions, each ion is in a separate pure quantum state.  
After the collision, the detected particles may be in a mixed state or even remain in a pure state depending on the number of particles detected (quantum entanglement) \cite{Sakurai:2011zz}.
During intermediate times when the particle density is high, scatterings may also play a role of quantum decoherence.
The hot and dense medium created is often modeled with hydrodynamics which assumes local thermal equilibrium and quantum decoherence above the thermal length scale.
Given the success of hydrodynamic modeling of high-energy collisions, it is conceptually difficult to imagine the survival of coherence in the final state.

The effect of coherence manifests itself most directly in the suppression of Bose-Einstein correlations \cite{Goldhaber:1960sf,Kopylov:1975rp,Gyulassy:1979yi,Andreev:1992pu,Akkelin:2001nd}.  
However, even for a coherent fraction as large as $50\%$, the intercept (vanishing relative momentum) of 2-pion correlations is decreased only by $25\%$. 
Furthermore, Coulomb repulsion and the dilution from long-lived resonance decays also suppress Bose-Einstein correlations.
Given the various sources of suppression and the unknown functional form of Bose-Einstein correlations, it is practically impossible to determine the coherent fraction from 2-pion correlations alone.

In this article, a set of techniques for the measurement of 3- and 4-pion QS correlations is presented.  
In particular, it is shown how the comparison of 2-, 3-, and 4-pion QS correlations can allow for a less ambiguous determination of the coherent fraction.  
To that purpose, the method of constructed or {\it built} correlation functions is introduced.
The techniques are presented for charged pions up to $4^{\rm th}$ order but can be easily generalized beyond this.

This article is organized into 10 sections.
In Sec.~\Rmnum{2}, the correlation functions and projection variables are defined.
In Sec.~\Rmnum{3}, the concept of symmetrization and pair exchange amplitudes is introduced.
The main technique of built correlation functions is introduced in Sec.~\Rmnum{4}.
A short discussion of radii measurements using multi-pion correlations is given in Sec.~\Rmnum{5}.
Final-state-interactions are discussed in Sec.~\Rmnum{6}.
Dilution effects from long-lived emitters are discussed in Sec.~\Rmnum{7}.
In Sec.~\Rmnum{8}, the cumulant and partial cumulant correlation functions are defined.
A variety of model calculations are shown in Sec.~\Rmnum{9}.
In Sec.~\Rmnum{10}, the effect of multi-boson distortions on the comparison of built and measured correlation functions is calculated.  
Finally, the main findings of this analysis are presented in Sec.~\Rmnum{11}.

\section{Distributions and correlation functions}
The $n$-particle inclusive momentum spectrum is given by $N_n(p_1,p_2,...,p_n)$ where $p_i$ denotes the momentum of particle $i$.
The product of $n$ single-particle inclusive spectra, $N_1(p_1)N_1(p_2)...N_1(p_n)$, forms a combinatorial reference sample to the full $n$-particle spectrum.
Correlation functions of order $n$ can then be constructed in the usual manner as the ratio of the two distributions,
\begin{equation}
C_{n}(p_1,p_2,...,p_n) = \frac{N_n(p_1,p_2,...,p_n)}{N_1(p_1)N_1(p_2)...N_1(p_n)}. \label{eq:Cn}
\end{equation}
Experimentally, the two distributions can be measured by taking all $n$ particles from the same collision event or by taking all $n$ from different but similar events.
In quantum optics, the maximum of $C_n$ for identical pions is $n!$ in the dilute gas limit \cite{Zajc:1986sq,Pratt:1993uy,Pratt:1994cg,Lednicky:1999xz,Zhang:1998sz} and neglecting isospin \cite{Akkelin:2001nd}.

\subsection{Projection variables}
Two-particle correlations are projected onto relative momentum in the Pair-Rest-Frame (PRF) or in the Longitudinally Co-Moving System (LCMS) where the z-component of the pair momentum vanishes.
In the PRF, the 1D Lorentz invariant relative momentum, $\qinv = \sqrt{-(p_1-p_2)^{\mu}(p_1-p_2)_{\mu}}$, is used.
In the LCMS frame, the relative momentum vector is projected onto 3 dimensions. 
The projection onto the pair momentum vector forms $q_{\rm out}$.
The projection onto the longitudinal direction (beam-axis) forms $q_{\rm long}$.
The direction perpendicular to both ``out'' and ``long'' forms the $q_{\rm side}$ projection.

For 3- and 4-particle correlations, the Lorentz invariant relative momentum is used and defined by the quadrature sum of pair invariant relative momenta 
\begin{eqnarray}
Q_3 &=& \sqrt{q_{12}^2+q_{13}^2+q_{23}^2}, \\
Q_4 &=& \sqrt{q_{12}^2+q_{13}^2+q_{14}^2+q_{23}^2+q_{24}^2+q_{34}^2},
\end{eqnarray}
for 3- and 4-particle correlations, respectively.
The transverse momentum dependence can be studied by also projecting onto the average transverse momenta
\begin{eqnarray}
\kT &=& \frac{|\vec{p}_{\rm T,1}+\vec{p}_{\rm T,2}|}{2}, \\
\KTThree &=& \frac{|\vec{p}_{\rm T,1}+\vec{p}_{\rm T,2}+\vec{p}_{\rm T,3}|}{3}, \\
\KTFour &=& \frac{|\vec{p}_{\rm T,1}+\vec{p}_{\rm T,2}+\vec{p}_{\rm T,3}+\vec{p}_{\rm T,4}|}{4},
\end{eqnarray}
for 2-, 3-, and 4-particle correlations, respectively.

\section{Symmetrization}

For identical pions, the effect of quantum statistics is represented as a symmetrization of pion creation points. 
The symmetrization is only valid for independently produced pions \cite{Goldhaber:1960sf,Kopylov:1975rp}. 
As pions in a coherent state are collectively produced, the symmetrization is absent for pairs of coherent pions.
It is, however, present for mixed-pairs with one pion from the coherent pool and the other from chaotic pool of particles.

\begin{figure}
\center
  \includegraphics[width=0.24\textwidth]{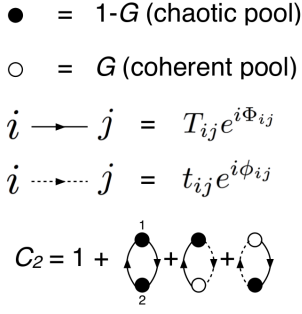}
  \caption{2-pion symmetrization diagrams.  Derived from similar figures in Ref.~\cite{Csorgo:1999sj}.}
  \label{fig:Symm2pion}
\end{figure}

In the Wigner function formalism and neglecting particle interactions, the correlation functions are written in terms of the Fourier transformed single-particle emission function, $S(x,p)$, which describes the phase-space distribution of particle production \cite{Csorgo:1994in}.
For the case of chaotic + coherent emission, the emission function is split into the sum of a chaotic and coherent parts \cite{Andreev:1992pu,Heinz:1997mr,Akkelin:2001nd}
\begin{equation}
 S(x,p) = [1-G(p)]S_{\rm ch}(x,p) + G(p)S_{\rm coh}(x,p).
\end{equation}
The fraction of pions from the coherent pool is given by $G(p)$ and is momentum dependent in general although it is treated as momentum independent for the rest of this article. 
The normalized pair exchange amplitudes, $d_{ij}$, are introduced in order to write the correlation functions compactly in terms of the emission functions
{\footnotesize
\begin{eqnarray}
d_{ij} &=& \frac{ \int d^4x S(x,K_{ij}) e^{iq_{ij}x} }{ [\int d^4x S(x,p_i) \int d^4y S(y,p_j)]^{1/2} } \nonumber \\
&=& \frac{ \int d^4x \left[(1-G)S_{\rm ch}(x,K_{ij})+GS_{\rm coh}(x,K_{ij})\right] e^{iq_{ij}x} }{ [\int d^4x S(x,p_i) \int d^4y S(y,p_j)]^{1/2} } \nonumber \\
&=& (1-G)T_{ij}e^{i\Phi_{ij}} + Gt_{ij}e^{i\phi_{ij}} \label{eq:EAgeneral}.
\end{eqnarray}}
The magnitudes of the Fourier transform (FT), $T_{ij}$ and $t_{ij}$, are referred to as the {\it pair exchange magnitudes} for the chaotic and coherent parts of the source, respectively.
The phases of the FT for the chaotic and coherent parts are given by $\Phi_{ij}$ and $\phi_{ij}$, respectively.
The average and relative pair momentum are given by $K_{ij}=(p_i+p_j)/2$ and $q_{ij}=p_i-p_j$, respectively.
The two-pion correlation function can then be written as in Ref.~\cite{Andreev:1992pu}
{\footnotesize
\begin{eqnarray}
&C_2(p_1,p_2) = 1& \\
&+ \frac{ |\int d^4x S(x,K_{12}) e^{iq_{12}x}|^2 - G^2|\int d^4x S_{\rm coh}(x,K_{12}) e^{iq_{12}x}|^2}{\int d^4x S(x,p_1)\int d^4y S(y,p_2)}& \\
&= 1 + (1-G)^2 T_{12}^2 + 2G(1-G)T_{12}t_{12}\cos(\Phi_{12}-\phi_{12})&  \label{eq:EAequation2}
\end{eqnarray}
}
From Eq.~\ref{eq:EAequation2}, one notices that while the FT phases ($\Phi,\phi$) disappear in the case of fully chaotic emission ($G=0$), they are present in the case of partial coherence.
Each term in Eq.~\ref{eq:EAequation2} can be represented in a symmetrization diagram similar to Ref.~\cite{Csorgo:1999sj} and is shown in Fig.~\ref{fig:Symm2pion}.
Pions from the chaotic pool are represented with a solid circle while those from the coherent pool are given by a hollow circle.
Chaotic pions yield a multiplicative factor of $1-G$ while coherent pions yield a factor of $G$.
Solid and dotted lines yield a multiplicative factor from the pair exchange amplitude for the chaotic ($T_{ij}e^{i\Phi_{ij}}$) and coherent ($t_{ij}e^{i\phi_{ij}}$) part of the source, respectively.
Lines running from pion $j$ to pion $i$ are the complex conjugate of those in the opposite direction.
The sum of all graphs yield the full QS correlation function.
Also, from Eq.~\ref{eq:EAgeneral} one has the equality, $\Phi_{ij}=-\Phi_{ji}$ and $\phi_{ij}=-\phi_{ji}$.

\begin{figure}
\center
  \includegraphics[width=0.37\textwidth]{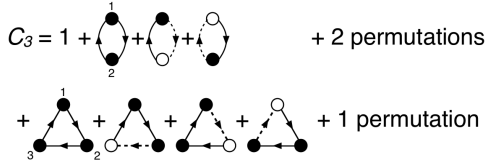}
  \caption{3-pion symmetrization diagrams.  Derived from similar figures in Ref.~\cite{Csorgo:1999sj}.}
  \label{fig:Symm3pion}
\end{figure}

\begin{figure}
\center
  \includegraphics[width=0.49\textwidth]{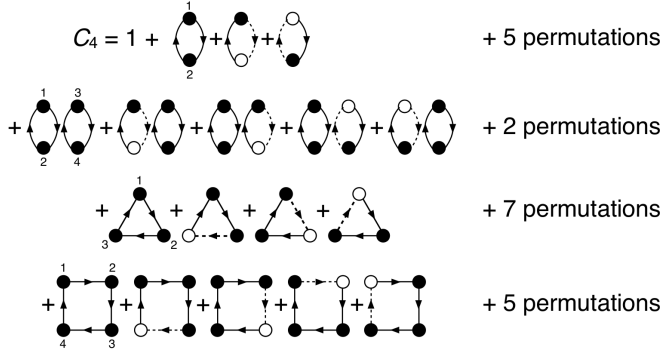}
  \caption{4-pion symmetrization diagrams.  Derived from similar figures in Ref.~\cite{Csorgo:1999sj}.}
  \label{fig:Symm4pion}
\end{figure}

The symmetrization diagrams for 3- and 4-pion correlations are shown in Fig.~\ref{fig:Symm3pion} and \ref{fig:Symm4pion}, respectively.
The number of permutations in the $n$-pion set is $n!$.
From the diagrams in Fig.~\ref{fig:Symm3pion}, the 3-pion correlation function can then be written as in Ref.~\cite{Andreev:1992pu}
{\footnotesize
\begin{eqnarray}
C_3(p_1,p_2,p_3) &=& 1 + (1-G)^2(T_{12}^2 + T_{23}^2 + T_{31}^2) \nonumber \\
&+& 2G(1-G)[T_{12}t_{12}\cos(\Phi_{12}-\phi_{12}) \nonumber \\
&+& T_{23}t_{23}\cos(\Phi_{23}-\phi_{23}) \nonumber \\
&+& T_{31}t_{31}\cos(\Phi_{31}-\phi_{31})] \nonumber \\
&+& 2G(1-G)^2 [T_{12}T_{23}t_{31}\cos(\Phi_{12}+\Phi_{23}+\phi_{31}) \nonumber \\
&+& T_{12}t_{23}T_{31}\cos(\Phi_{12}+\phi_{23}+\Phi_{31}) \nonumber \\
&+& t_{12}T_{23}T_{31}\cos(\phi_{12}+\Phi_{23}+\Phi_{31})] \nonumber \\
&+& 2(1-G)^3 T_{12}T_{23}T_{31}\cos(\Phi_{12}+\Phi_{23}+\Phi_{31}) \label{eq:C3full}.
\end{eqnarray}
}

\subsection{$r_3$ and $r_4$}
In the absence of pion coherence ($G=0$), the 3-pion FT phase factor in Eq.~\ref{eq:C3full} ($\cos(\Phi_{12}+\Phi_{23}+\Phi_{31})$) may be isolated by comparing 3-pion cumulant to 2-pion QS correlations.  
The 3-pion QS cumulant correlation has, by definition, all 2-pion symmetrization terms removed from Eq.~\ref{eq:C3full} and is therefore given by 
{\footnotesize
\begin{equation}
{\rm {\bf c}}_3(1,2,3) = 1 + 2T_{12}T_{23}T_{31}\cos(\Phi_{12}+\Phi_{23}+\Phi_{31}), \label{eq:c3EA}
\end{equation}}
where the simplified notation $(i)$ stands for $p_i$.
One may isolate the 3-pion phase factor by normalizing ${\rm {\bf c}}_3$ with the appropriate 2-pion correlation factors given in Eq.~\ref{eq:EAequation2} \cite{Heinz:1997mr,Heinz:2004pv}:
{\footnotesize
\begin{eqnarray}
r_3(1,2,3) &=& \frac{{\rm {\bf c}}_3(1,2,3)-1}{\sqrt{(C_2(1,2)-1)(C_2(2,3)-1)(C_2(3,1)-1)}} \nonumber \\
&=& 2\cos(\Phi_{12}+\Phi_{23}+\Phi_{31}) \label{eq:r3}.
\end{eqnarray}
}
In Ref.~\cite{Heinz:1997mr} it was shown that to leading order, the relative momentum dependence of $r_3$ is quartic.
In the absence of the 3-pion phase, $r_3$ is equal to 2.0 for all triplet relative momenta.
The $r_3$ function was recently measured at the LHC \cite{Abelev:2013pqa} in Pb-Pb collisions.  
Although the systematic uncertainties were quite large for high $Q_3$, no significant $Q_3$ dependence was observed for both low and high transverse momentum.

Similarly, the 4-pion FT phase factor ($\cos(\Phi_{12}+\Phi_{23}+\Phi_{34}+\Phi_{41})$) may be isolated by comparing 4-pion cumulant to 2-pion QS correlations.  
The 4-pion QS cumulant correlation has all 2-pion, 2-pair, and 3-pion symmetrization terms explicitly removed and is therefore given by 
{\footnotesize
\begin{eqnarray}
&{\rm {\bf c}}_4(1,2,3,4) = 1 + 2T_{12}T_{23}T_{34}T_{41}\cos(\Phi_{12}+\Phi_{23}+\Phi_{34}+\Phi_{41})& \nonumber \\
&+ 2T_{12}T_{24}T_{43}T_{31}\cos(\Phi_{12}+\Phi_{24}+\Phi_{43}+\Phi_{31})& \nonumber \\
&+ 2T_{13}T_{32}T_{24}T_{41}\cos(\Phi_{13}+\Phi_{32}+\Phi_{24}+\Phi_{41})& \nonumber \\
&\rightarrow 1 + 6T_{12}T_{23}T_{34}T_{41}\cos(\Phi_{12}+\Phi_{23}+\Phi_{34}+\Phi_{41}),& \label{eq:c4EA}
\end{eqnarray}
}
where the last equality is naturally obtained after averaging over many pion quadruplets.
The 4-pion FT phase can then be isolated with $r_4$ defined as
{\footnotesize
\begin{eqnarray}
r_4(1,2,3,4) &=& \frac{{\rm {\bf c}}_4(1,2,3,4)-1}{\sqrt{(C_2(1,2)-1)(C_2(2,3)-1)(C_2(3,4)-1)(C_2(4,1)-1)}} \nonumber \\
&=& 6\cos(\Phi_{12}+\Phi_{23}+\Phi_{34}+\Phi_{41}) \label{eq:r4}.
\end{eqnarray}
}

\section{Building multi-pion QS correlation functions}
In the absence of multi-pion FT phases and coherence, higher order correlation functions ($n>2$) do not contain any additional information beyond that which is already present in 2-pion correlation functions.
The absence/presence of both phenomena may then be tested by comparing the measured multi-pion QS correlations to the expectations from measured 2-pion correlations.
For the rest of this section the FT phases ($\Phi_{ij}=0,\phi_{ij}=0$) are neglected as one expects a rather weak relative momentum dependence \cite{Heinz:1997mr} and since the recent ALICE $r_3$ results show no substantial $Q_3$ dependence \cite{Abelev:2013pqa}.
With this simplification the building blocks of higher order correlation functions are the pair exchange magnitudes, $T_{ij}, t_{ij}$, and the coherent fraction, $G$.

From Figs.~\ref{fig:Symm2pion}-\ref{fig:Symm4pion}, 2-, 3-, and 4-pion QS correlations can be written as in Ref.~\cite{Andreev:1992pu}:
{\footnotesize
\begin{eqnarray}
C_2(1,2) &=& 1 + (1-G)^2 T_{12}^2 + 2G(1-G)T_{12}t_{12} \label{eq:C2NoPhase} \\
C_3(1,2,3) &=& 1 + (1-G)^2(T_{12}^2 + T_{23}^2 + T_{31}^2) \nonumber \\
&+& 2G(1-G)[T_{12}t_{12} + T_{23}t_{23} + T_{31}t_{31}] \nonumber \\
&+& 2G(1-G)^2 [T_{12}T_{23}t_{31} + T_{12}t_{23}T_{31} + t_{12}T_{23}T_{31}] \nonumber \\
&+& 2(1-G)^3 T_{12}T_{23}T_{31} \label{eq:C3NoPhase} \\
C_4(1,2,3,4) &=& 1 + (1-G)^2(T_{12}^2 + T_{23}^2 + T_{31}^2 + T_{14}^2 + T_{24}^2 + T_{34}^2) \nonumber \\ 
&+& 2G(1-G)[T_{12}t_{12} + T_{23}t_{23} + T_{31}t_{31} \nonumber \\ 
&+& \mlq14\mrq + \mlq24\mrq + \mlq34\mrq] \nonumber \\ 
&+& (1-G)^4[T_{12}^2T_{34}^2 + T_{31}^2T_{24}^2 + T_{14}^2T_{23}^2] \nonumber \\ 
&+& G(1-G)^3[T_{12}^2T_{34}t_{34} + T_{34}^2T_{12}t_{12} \nonumber \\
&+& \mlq13,24\mrq + \mlq14,23\mrq] \nonumber \\
&+& 2(1-G)^3 [T_{12}T_{23}T_{31} + T_{12}T_{24}T_{14} \nonumber \\ 
&+&  T_{31}T_{34}T_{41} + T_{23}T_{34}T_{42}]\nonumber \\ 
&+& 2G(1-G)^2 [T_{12}T_{23}t_{31} + T_{12}t_{23}T_{31} + t_{12}T_{23}T_{31} \nonumber \\
&+& \mlq124\mrq + \mlq134\mrq + \mlq234\mrq] \nonumber \\
&+& 2(1-G)^4 [T_{12}T_{23}T_{34}T_{41} \nonumber \\ 
&+& T_{12}T_{24}T_{43}T_{31} + T_{13}T_{32}T_{24}T_{41}] \nonumber \\
&+& 2G(1-G)^3 [T_{12}T_{23}T_{34}t_{14} + T_{12}T_{23}t_{34}T_{14} \nonumber \\
&+& T_{12}t_{23}T_{34}T_{14} + t_{12}T_{23}T_{34}T_{14} \nonumber \\
&+& \mlq4321\mrq + \mlq1243\mrq + \mlq3421\mrq + \mlq1324\mrq + \mlq4231\mrq], \label{eq:C4NoPhase}
\end{eqnarray}
}
where $\mlq ij \mrq$, $\mlq ij,kl \mrq$, $\mlq ijk \mrq$, and $\mlq ijkl \mrq$ represent permutations within the same symmetrization sequence.
In the case of fully chaotic emission, $T_{ij}$ can be extracted directly from 2-pion correlations.
As $T_{ij}$ is a 6D function, experimental 2-pion correlations should be binned as differentially as possible.
One may exploit longitudinal boost invariance and bin in the LCMS in 4D ($\kT,q_{\rm out},q_{\rm side},q_{\rm long}$) as done in Ref.~\cite{Abelev:2013pqa}.

In the case of partial coherence one has three quantities which cannot be extracted from 2-pion correlations alone: $G,T_{ij},t_{ij}$.
One can, however, consider two extreme scenarios of coherent emission.
In one case it is assumed that the space-time structure of coherent emission is identical to chaotic emission ($t_{ij}=T_{ij}$).
In another case one assumes that coherent emission is point-like at the center of the collision which might be expected for Bose-Einstein condensate ($t_{ij}=1$).
One may also consider additional scenarios with a suitable parametrization of $t_{ij}$ (e.g.\ Gaussian structure with an assumed radius).
The chaotic pair exchange magnitude, $T_{ij}$, can then be extracted from Eq.~\ref{eq:C2NoPhase} with various assumptions of $G$.
For the purpose of studying partial coherence, {\it built} multi-pion correlation functions are defined with Eqs.~\ref{eq:C3NoPhase}-\ref{eq:C4NoPhase} but with $T_{ij}$ taken from lower order measurements.
A minimization of the $\chi^2$ between built and measured correlations for each $Q_n$ bin can be used to estimate $G$ for different assumptions of the coherent source profile.

\subsection{Extraction of $T_{ij}$ from ${\rm {\bf c}}_3$}
In high-multiplicity collision events, femtoscopy lies in a clean region of phase-space where background correlations unrelated to QS and final-state-interactions (FSI) are negligible.  
However, in low-multiplicity events, background correlations such as mini-jets \cite{Aamodt:2010jj,Aamodt:2011kd} are non-negligible and not exactly known.
The extraction of $T_{ij}$ from 2-pion correlations can then be unreliable.
Instead, one may exploit 3-pion cumulants as was done in Ref.~\cite{Abelev:2014pja} for which 2-pion background correlations are explicitly removed.
The 3-pion cumulant can be binned in 3D pair invariant relative momenta ($q_{12},q_{23},q_{31}$).  
The pair exchange magnitude can be parametrized with an Edgeworth or Laguerre expansion \cite{Csorgo:2000pf}
\begin{eqnarray}
T_{ij} &=& sE_{\rm w}(R\,q_{ij})\,e^{-R^2\,q_{ij}^2/2} \label{eq:EWEA} \\
E_{\rm w}(R,q_{ij}) &=& 1 + \sum_{n=3}^{\infty} \frac{\kappa_n}{n! (\sqrt{2})^n} H_n(R\,q_{ij})\,, \\
T_{ij} &=& sL_{\rm g}(R\,q_{ij})\,e^{-R\,q_{ij}/2} \\
L_{\rm g}(R\,q_{ij}) &=& 1 + \sum_{n=1}^{\infty} \frac{l_n}{n!} L_n(R\,q_{ij})\,.
\end{eqnarray}
For the Edgeworth expansion, $E_{\rm w}(Rq_{ij})$ characterizes deviations from Gaussian behavior, 
$H_n$ are the Hermite polynomials, and $\kappa_n$ are the Edgeworth coefficients.  
For the Laguerre expansion, $L_{\rm g}(Rq_{ij})$ characterizes deviations from exponential behavior, 
$L_n$ are the Laguerre polynomials, and $l_n$ are the Laguerre coefficients.  
In both expansions, $s$ is an additional scale parameter and $R$ is the $2^{\rm nd}$ cumulant of the correlation function.
For the case when $E_{\rm w}=1$, $R$ is the standard deviation of a Gaussian source profile. 
For the case when $L_{\rm g}=1$, $R$ is the FWHM of a Cauchy (Lorentzian) source profile. 

The 3-pion cumulant correlation can then be fit according to Eq.~\ref{eq:c3EA} to extract the coefficients of the $T_{ij}$ parametrization.
The resulting $T_{ij}$ can be used to build correlation functions.
However, since ${\rm {\bf c}}_3(q_{12},q_{23},q_{31})$ is only a 3D projection of a 9D correlation function, there are somewhat more limitations to the accuracy of $T_{ij}$ extracted from 3-pion cumulants than 2-pion correlations. 
An estimate for the bias on built correlation functions caused by limited dimensionality and binning will be discussed in section X.

\section{Radii measurements}
Most often in experimental analyses, 2-pion QS correlations are used to extract information on the freeze-out space-time structure of the particle emitting source (e.g.\ radii).
Less explored are the multi-pion QS correlations which may also be used to extract freeze-out radii \cite{Ackerstaff:1998py,Abreu:1995sq,Achard:2002ja,Abelev:2014pja}.
For such measurements, one typically assumes fully chaotic emission and parametrizes $T_{ij}$ in Eqs.~\ref{eq:C3NoPhase}-\ref{eq:C4NoPhase} with the usual Gaussian or exponential forms.

Concerning the dimensionality of multi-pion projections, often the correlations are projected and fit with 1D Lorentz invariant relative momentum, $Q_n$.
However, a 1D fit to the full correlation function is incorrect due to the summation of terms with different powers of $T_{ij}$ in Eqs.~\ref{eq:C3NoPhase} and \ref{eq:C4NoPhase}. 
Instead, one may isolate and fit only the 3-pion cumulant, ${\rm {\bf c}}_3$, for which the 2-pion symmetrization terms are explicitly removed.  
The cumulant may be fit in 1D with a Gaussian function only since the definition of $Q_3$ coincides with the Gaussian relative momentum dependence of $T_{12}T_{23}T_{31}$.
For a spherically symmetric Gaussian source of radius $R$, $T_{ij}=e^{-R^2q_{ij}^2/2}$ and therefore $T_{12}T_{23}T_{31}=e^{-R^2Q_3^2/2}$.
For all other fit functions, it is more appropriate to fit in 3D (${\rm {\bf c}}_3(q_{12},q_{23},q_{31})$).
In the case of 4-pion correlations, even a Gaussian fit to the cumulant correlation function is not correct since $Q_4$ does not coincide with the relative momentum dependence of $T_{ij}T_{jk}T_{kl}T_{li}$.
The source radius may be measured from ${\rm {\bf c}}_4$ through the 6D pair $q$ information by performing a fit according to Eq.~\ref{eq:c4EA} with the appropriate parametrization of the $T$ factors.
To avoid such high dimensional histograms, one may also compute Eq.~\ref{eq:c4EA} for each quadruplet (online) for a variety of $R$ choices and project against $Q_4$.
One may then perform a $\chi^2$ minimization offline to determine the best fit $R$ value.

\subsection{Effect of partial coherence on extracted radii}
\begin{figure}
\center
  \includegraphics[width=0.49\textwidth]{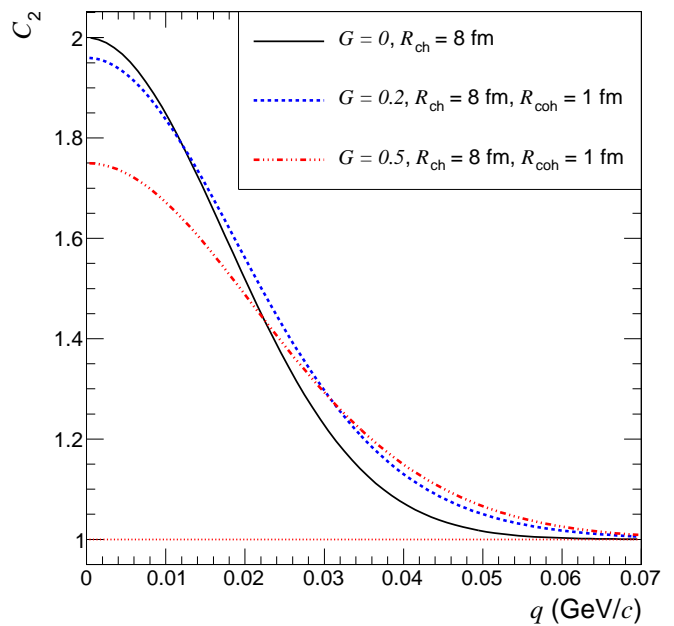}
  \caption{(Color online) Modification of 2-pion QS correlation functions in the presence of partial coherence.  The correlation function for a fully chaotic spherically symmetric Gaussian source with $R_{\rm ch}=8$ fm is shown with a solid black line.  Dashed blue and dash-dotted red lines include an additional spherically symmetric coherent component with $R_{\rm coh}=1$ fm for $G=0.2$ and $G=0.5$, respectively.}
  \label{fig:C2width}
\end{figure}

The extraction of femtoscopic radii in high-energy collision data is universally done with a fully chaotic assumption of particle emission.
It is well known, however, that the presence of partial coherence does not only suppress the intercept of QS correlations but may also modify its width \cite{Gyulassy:1979yi,Andreev:1992pu,Akkelin:2001nd}.  
With partial coherence in Eq.~\ref{eq:EAequation2}, one may expect the same extracted radius only when $T_{ij}=t_{ij}$ and $\Phi_{ij}=\phi_{ij}$ (i.e.\ identical space-time structure of both chaotic and coherent components).
Furthermore, the coherent fraction must be momentum independent.
In the more probable cases where these conditions are not satisfied, partial coherence will modify the usual mapping of the correlation function width to the freeze-out radius.
To illustrate this effect, 2-pion correlation functions are compared without coherence to two cases with partial coherence in Fig.~\ref{fig:C2width}.
One observes that even with a substantial coherent fraction of $20\%$, the suppression at the intercept ($q=0$) is very small.  
Moreover, above $q=0.15$ MeV/$c$ there is actually an enhancement as compared to the fully chaotic case due to the quantum interference between the chaotic and the smaller coherent source (wider correlation).
Thus, in the presence of a smaller coherent component, the traditional fitting procedure will underestimate the chaotic source radius.
A single Gaussian fit to the red dash-dotted curve in Fig.~\ref{fig:C2width} yields a radius $25\%$ smaller than the chaotic source radius.

\section{Treatment of Final-State-Interactions}
The measurement of pure QS correlations is complicated by the presence of FSI (e.g.\ Coulomb repulsion).
Fortunately, the Coulomb+strong wave-functions for 2-pion correlations are well known to high accuracy \cite{Sinyukov:1998fc,Lednicky:2005tb},
\begin{eqnarray}
&\psi_{-{\bf k^*}}({\bf r^*}) = e^{i\delta_c}\sqrt{A_c(\eta)}& \nonumber \\
&\times \left[ e^{-i{\bf k^*r^*}}F(-i\eta,1,i\xi) + f_c(k^*)\frac{\tilde{G}(\rho,\eta)}{r^*} \right],& \label{eq:FSIWF}
\end{eqnarray}
where ${\bf k^*}$ and ${\bf r^*}$ are the momentum and relative separation evaluated in the PRF, $\delta_c$ is the Coulomb s-wave phase shift, $\eta=(k^*a)^{-1}$, $\xi={\bf k^*r^*} + k^*r^*$, $\rho=k^*r^*$, $A_c=2\pi\eta[e^{2\pi\eta}-1]^{-1}$ is the Gamov factor, $F$ is the confluent hypergeometric function, $f_c$ is the strong scattering amplitude re-normalized by the long-range Coulomb forces, $\tilde{G}$ is an s-wave Coulomb function, and $a$ is the Bohr radius taking into account the sign of the interaction.
The 2-pion FSI correlation can be computed by averaging the modulus square of the symmetrized wave-function over an assumed freeze-out source profile and dividing by the same average done with pure plane-waves
\begin{equation}
K_2(q) = \frac{\left<|\psi_{-{\bf k^*}}({\bf r^*}) + \psi_{{\bf k^*}}({\bf r^*})|^2\right>}{\left<|e^{-i{\bf k^*r^*}} + e^{i{\bf k^*r^*}}|^2\right>}. \label{eq:K2}
\end{equation}
Note that in the PRF, $q=2k^*$.
Often, the averaging is done with a spherical Gaussian source profile.  
A more accurate averaging would be one which includes the larger $r^*$ tails created by pions originating from resonance decays.
However, due to the relatively large value of the pion Bohr radius ($\pm 388$ fm for like/unlike sign pairs) as compared to the typical value of $r^*\sim20$ fm in high-energy heavy-ion collisions, the short-range structure of the source only mildly alters the functional form of $K_2(q)$. 
The averaging of $K_2(q)$ is usually confined to the core of particle production where $r^*<100$ fm.

\subsection{Multi-body FSI}
The exact wave-function of the $n$-body ($n>2$) Coulomb scattering is unknown.
However, solutions do exist in all asymptotic regions of phase-space \cite{Alt:1992js}.
In particular, the region of phase-space where all inter-particle spacings are sufficiently large is known as $\Omega_0$.
In the $\Omega_0$ asymptotic limit, the $n$-body Coulomb system is a sum of $\frac{n!}{(n-2)!2!}$ non-interacting two-body systems \cite{LednickyAmelin,Alt:1998nr}.
It has been shown to be highly successful in describing ionization by electron impact ($e,2e$) in atomic physics \cite{Alt:2005su}.
The relevant triplet kinetic energy for the applicability of the $\Omega_0$ ansatz was estimated in Ref.~\cite{Alt:1998nr} 
\begin{equation}
E_{\rm tot} \geq 0.2 \frac{\hbar c}{R({\rm fm})},
\end{equation}
where $R$ is the estimated source radius.
The characteristic Gaussian 1D radii measured in pp/p--Pb and central Pb--Pb collisions at the LHC are about 1.5 and 9 fm, respectively \cite{Abelev:2014pja}.
The resulting threshold triplet energies are therefore 26 and 4 MeV for pp/p--Pb and Pb--Pb, respectively. 
For a symmetric triangular configuration of momenta in the triplet rest frame, the corresponding threshold for $Q_3$ is identical to the $E_{tot}$ value.
Above the threshold, particles rapidly separate from each other such that the shorter range 3-body Coulomb forces (not treated in the $\Omega_0$ ansatz) become negligible.
At the 4-pion level, the $\Omega_0$ ansatz should remain a good approximation provided that each triplet $Q_3$ in the quadruplet is above the threshold.
The 3-pion symmetrized Coulomb wave-function in $\Omega_0$ is given as in Ref.~\cite{LednickyAmelin,Alt:1992js,Alt:1998nr} by
{\footnotesize
\begin{eqnarray}
\psi_{p_1,p_2,p_3}(x_1,x_2,x_3) = \nonumber \\
\frac{1}{\sqrt{6}} [ e^{i(q_{12}'x_{12} + q_{13}'x_{13} + q_{23}'x_{23})/3}\Phi_{12}(x_{12})\Phi_{13}(x_{13})\Phi_{23}(x_{23}) \nonumber \\
+ e^{i(q_{12}'x_{21}+q_{13}'x_{23}+q_{23}'x_{13})/3}\Phi_{12}(x_{21})\Phi_{13}(x_{23})\Phi_{23}(x_{13}) \nonumber \\ 
+ e^{i(q_{12}'x_{32}+q_{13}'x_{31}+q_{23}'x_{21})/3}\Phi_{12}(x_{32})\Phi_{13}(x_{31})\Phi_{23}(x_{21}) \nonumber \\ 
+ e^{i(q_{12}'x_{13}+q_{13}'x_{12}+q_{23}'x_{32})/3}\Phi_{12}(x_{13})\Phi_{13}(x_{12})\Phi_{23}(x_{32}) \nonumber \\
+ e^{i(q_{12}'x_{23}+q_{13}'x_{21}+q_{23}'x_{31})/3}\Phi_{12}(x_{23})\Phi_{13}(x_{21})\Phi_{23}(x_{31}) \nonumber \\
+ e^{i(q_{12}'x_{31}+q_{13}'x_{32}+q_{23}'x_{12})/3}\Phi_{12}(x_{31})\Phi_{13}(x_{32})\Phi_{23}(x_{12})] \label{eq:Psi3},
\end{eqnarray}}
where $q_{ij}'=p_i-p_j$, $x_{ij}=x_i-x_j$, and $\Phi_{ij}(x_{kl}) = \sqrt{A_c(\eta_{ij})}F(-i\eta_{ij},1,i({\bf k_{ij}^*r_{kl}^*} + k_{ij}^*r_{kl}^*))$ is the Coulomb modulation factor and is evaluated in the $ij$ PRF.
The 3-pion Coulomb correlation is then defined similar to $K_2$ as
\begin{equation}
K_3(q_{12},q_{13},q_{23}) = \frac{\left< |\psi_{p_1,p_2,p_3}(x_1,x_2,x_3)|^2 \right>}{\left< |\psi_{p_1,p_2,p_3}(x_1,x_2,x_3;\Phi=1)|^2 \right>}
\end{equation}

In practice, the calculation of $n$-body $\Omega_0$ wave-functions is quite involved.  
Moreover, its application to experimental data requires a multi-dimensional calculation in terms of the pair invariant relative momenta.
For Coulomb correlations of $4^{\rm th}$ order, a 6D calculation is required.  
As such a calculation is not very practical, a simplified approach is desirable.  

Another approach is the so-called generalized Riverside (GRS) method which treats the $n$-body FSI correlation as a product of 2-body FSI factors
{\footnotesize
\begin{eqnarray}
&K_3(q_{12},q_{13},q_{23}) = K_2(q_{12})K_2(q_{13})K_2(q_{23}),& \label{eq:K3GRS} \\
&K_4(q_{12},q_{13},q_{14},q_{23},q_{24},q_{34})& \nonumber \\
&= K_2(q_{12})K_2(q_{13})K_2(q_{14})K_2(q_{23})K_2(q_{24})K_2(q_{34}).& \label{eq:K4GRS} 
\end{eqnarray}}
The GRS approach easily allows a multi-dimensional estimate of multi-body FSI since only pair calculations are required.
Calculations of both approaches are presented in Sec.~\ref{sec:THERM}.

\section{Treatment of dilution from non-femtoscopic separations}
The isolation of pure QS correlations is further complicated by the contribution of pions from long-lived emitters.
Such pions are typically separated from other pions by many tens to hundreds of femtometers at freeze-out for which the expected quantum interference peak is only at extremely low $q$ ($<5$ MeV/$c$) \cite{Lednicky:1979ig}.
In the moderate $q$ region, pions from long-lived emitters effectively {\it dilute} the correlation function.  
Long-lived emitters include weak-decays (secondary contamination) as well as long-lived resonance decays such as the $\eta$. 

The treatment of the dilution for 2-pion correlations is usually done with the following formula connecting the measured and QS distributions/correlation functions \cite{Sinyukov:1998fc}:
{\footnotesize
\begin{eqnarray}
N_2(p_1,p_2) &=& (1-f_c^2)N_1(p_1)N_1(p_2) + f_c^2 K_2(q) N_2^{QS}(p_1,p_2), \label{eq:N2QS} \\
C_2(q) &=& (1-f_c^2) + f_c^2 K_2(q) C_2^{QS}(q). \label{eq:C2QS}
\end{eqnarray}}
The distribution, $N_2(p_1,p_2)$, can be experimentally formed by sampling both pions from the same event whereas $N_1(p_1)N_1(p_2)$ is sampled from separate events.
The correlated fraction of pairs for which an observable femtoscopic correlation (QS+FSI) exists is denoted by $f_c^2$. 
In the ``core/halo'' picture of particle production \cite{Csorgo:1994in}, particles may originate from one of two sources with very different characteristic radii \cite{Lednicky:1979ig}.  
Those originating from short-lived emitters are from the core of particle production and contain an observable QS correlation with other core particles.
Those from long-lived emitters are from the halo and do not observably interact with any other particle.
In such a picture, the fraction of particles from the core is denoted by $f_c$.

The treatment of dilutions in the core/halo picture can be easily extended to 3-pion correlations.
For the same-event 3-pion distribution, $N_3(p_1,p_2,p_3)$, there are four possible configurations.
In one case, all three pions originated from the core for which the probability is $f_c^3$.
In the second case, two pions originated from the core for which the probability is $3f_c^2(1-f_c)$.
In the third case, only one pion originated from the core with a probability of $3(1-f_c)^2f_c$.
In the final case, zero pions originated from the core with a probability of $(1-f_c)^3$.
The measured 3-pion same-event distribution can then be decomposed into its constituent parts while also utilizing Eq.~\ref{eq:N2QS} for the decomposition of $N_2(p_1,p_2)$ into its two parts \cite{Abelev:2013pqa}.
The same-event 3-pion distribution can be written as
{\footnotesize
\begin{eqnarray}
N_3(p_1,p_2,p_3) &=& f_{31} N_1(p_1)N_1(p_2)N_1(p_3) \nonumber \\
&+& f_{32} [N_2(p_1,p_2)N_1(p_3) + N_2(p_2,p_3)N_1(p_1) \nonumber \\
&+& N_2(p_3,p_1)N_1(p_2)] \nonumber \\
&+& f_{33} K_3 N_3^{QS}(p_1,p_2,p_3), \label{eq:N3} 
\end{eqnarray}}
where $N_3^{QS}$ is the quantity of interest to be extracted from the measured distributions (no QS superscript) and $K_3$ is the 3-body FSI factor.
The additional distribution, $N_2(p_i,p_j)N_1(p_k)$, provides information from 2-pion symmetrizations alone in the dilute gas limit.  
It may be experimentally measured by taking two pions from a single-event while taking the third from a mixed-event.
In the core/halo picture, $f_{31}=(1-f_c)^3 + 3f_c(1-f_c)^2 - 3(1-f_c)(1-f_c^2)$, $f_{32} = 1-f_c$, and $f_{33} = f_c^3$.

Four-pion correlations have four sequences of symmetrizations as illustrated in Fig.~\ref{fig:Symm4pion}. 
Isolation of each sequence can be accomplished with the following additional distributions:
\begin{eqnarray}
&N_1(p_1)N_1(p_2)N_1(p_3)N_1(p_4),& \\
&N_2(p_1,p_2)N_1(p_3)N_1(p_4),& \\
&N_2(p_1,p_2)N_2(p_3,p_4),& \label{eq:TwoPairDist} \\
&N_3(p_1,p_2,p_3)N_1(p_4),& \\
&N_4(p_1,p_2,p_3,p_4),&
\end{eqnarray}
which can be experimentally measured by taking the appropriate number of particles from the same event and the rest from mixed events.
For Eq.~\ref{eq:TwoPairDist}, one takes two particles from one event and the other two from a separate event.
Extending the treatment of dilutions to the 4-pion case, one obtains
{\footnotesize
\begin{eqnarray}
N_4(p_1,p_2,p_3,p_4) &=& f_{41} N_1(p_1)N_1(p_2)N_1(p_3)N_1(p_4) \nonumber \\
&+& f_{42} [N_2(p_1,p_2)N_1(p_3)N_1(p_4) \nonumber \\
&+& \mlq13\mrq + \mlq14\mrq + \mlq23\mrq + \mlq24\mrq +  \mlq34\mrq]\nonumber \\
&+& f_{43} [N_3(p_1,p_2,p_3)N_1(p_4) \nonumber \\
&+& \mlq124\mrq + \mlq134\mrq + \mlq234\mrq]\nonumber \\
&+& f_{44} K_4 N_4^{QS}(p_1,p_2,p_3,p_4), \label{eq:N4} 
\end{eqnarray}}
where $N_4^{QS}$ is again the quantity of interest to be extracted from the measured distributions and the 4-body FSI correlation, $K_4$.
Permutations indicated by $\mlq ij\mrq$ represent $N_2(p_i,p_j)N_1(p_k)N_1(p_l)$.
Permutations indicated by $\mlq ijk\mrq$ represent $N_3(p_i,p_j,p_k)N_1(p_l)$.
In the core/halo picture, $f_{41} = -3(1-f_c)^4 - 8f_c(1-f_c)^3 + 6(1-f_c^2)(1-f_c)^2$, $f_{42} = -(1-f_c)^2$, $f_{43} = (1-f_c)$, $f_{44}=f_c^4$.
Another possibility not represented in Eq.~\ref{eq:N4} is given by two pairs of interacting pions, $N_2(p_1,p_2)N_2(p_3,p_4)$.  
This possibility is absent in the core/halo picture and will be shown to be quite small in a more realistic model in Sec.~\ref{sec:THERM}.

\subsection{$\lambda$ parameter}
Most often in experimental analyses, the parameter $f_c^2$ in Eq.~\ref{eq:C2QS} is replaced with the so-called $\lambda$ parameter and $C_2^{QS}$ is parametrized without coherence as: $C_2^{QS} = 1 + e^{-(qR)^\alpha}$ \cite{Csorgo:2003uv,Adams:2004yc,Aamodt:2011mr}.
Such a parametrization can be flawed in regards to FSI corrections ($\lambda K_2$) in two noteworthy ways.  
First, when the true functional form of $C_2^{QS}$ is not known, a mismatch between the true and assumed function form can cause a large bias on the extracted $\lambda$ parameter \cite{Abelev:2013pqa}.
As the FSI correction is the product of $\lambda$ and $K_2$, a biased value of $\lambda$ also biases the FSI correction.
Second, coherent emission leads to the suppression of $C_2^{QS}$ but not of FSI correlations.  
In both cases, a single suppression parameter ($\lambda$) is not appropriate.
The problem can be circumvented by parametrizing the QS correlations instead by $C_2^{QS} = 1 + \lambda' e^{-(qR)^\alpha}$, where in addition to $f_c^2$ in Eq.~\ref{eq:C2QS}, a second suppression parameter is used.  

As in Ref.~\cite{Abelev:2013pqa}, $f_c^2$ may be determined less ambiguously from a fit to $\pi^+\pi^-$ correlation functions for which only FSI correlations contribute.
Assuming the equivalence of $f_c^+$ with $f_c^-$ as expected at high energies \cite{Akkelin:2001nd}, $\pi^+\pi^-$ correlations can be fit according to $C_2^{+-} = (1-f_c^2) + f_c^2 K_2(q,R)$.
Owing to the large value of the pion Bohr radius (388 fm) as compared to the typical relative separation at freeze-out in high-energy heavy-ion collisions ($\sim 20$ fm), $\pi^+\pi^-$ FSI correlations are much less sensitive to the short-range structure of the source.
A fit to $\pi^+\pi^-$ correlations is then simpler in practice as $K_2(q,R)$ changes less rapidly for different source profiles with the same characteristic radius.

The chaotic upper limit to the conventional $\lambda$ parameter is unity.  
However, experiments overwhelmingly report values less than unity--sometimes as low as 0.3 \cite{Adams:2004yc}.
There can be several sources of suppression for the conventional $\lambda$ parameter which are listed below in approximate order of magnitude:
\begin{itemize}
\item Dilution from pions of long-lived emitters
\item Gaussian fits to non-Gaussian correlations
\item Detector momentum smearing
\item Pion mis-identification
\item Finite $q$ bin width
\end{itemize}
An additional more interesting source of suppression not listed above, is the possibility of pion coherence.  However, the suppression is expected to be rather small for 2-pion correlations which was estimated to be $\sim 0.05$ at the LHC \cite{Abelev:2013pqa}.
The exact order of the above contributions depends on the experimental conditions of the analysis.
One of the main reasons for low values of $\lambda$ is often due to the use of Gaussian fits to correlation functions which are intrinsically non-Gaussian.
In Ref.~\cite{Abelev:2013pqa}, an estimate can be made for the top three sources of suppression.
A Gaussian fit to $C_2^{++}(q)$ yielded $\lambda \sim 0.4$.  
However, a fit to $C_2^{+-}(q)$ gave $\lambda \sim 0.7$, which indicates a suppression of about 0.3 from long-lived emitters in ALICE.
The separate analysis of $r_3$ revealed a possible coherent fraction of about $23\%$ for which the suppression to the intercept of 2-pion correlations is $0.23^2=0.05$.
This leaves a remaining suppression of about 0.25 due to non-Gaussian features in the correlation function.

Note that the non-Gaussian features for a 1D analysis are expected to differ from 3D analyses ($q_{\rm out},q_{\rm side},q_{\rm long}$) even if the source is Gaussian in all three dimensions but with different radii in each dimension.
The STAR collaboration reported values between 0.3 and 0.4 for 3D Gaussian fits \cite{Adams:2004yc} which likely indicates substantial non-Gaussian features even in 3D heavy-ion analyses.

\section{cumulant and partial cumulant correlation functions}
The full $n$-pion correlation function contains the full set of symmetrizations as previously illustrated.  
One may isolate different levels of symmetrization by subtracting different types of $n$-pion spectra.
The 3-pion cumulant correlation function can be defined as
\begin{eqnarray}
{\rm {\bf c}}_3 &=& [N_3^{\rm QS}(p_1,p_2,p_3) - \delta_1N_2^{\rm QS}(p_1,p_2)N_1(p_3) \nonumber \\
&+& \delta_2N_1^3] / N_1^3. \label{eq:c3}
\end{eqnarray}
where $N_1^3$ denotes $N_1(p_1)N_1(p_2)N_1(p_3)$ for brevity.
For the case of same-charge triplets, $\delta_1=\delta_2=3$.  
For the mixed-charge case, $\delta_1=\delta_2=1$.
The cumulant correlation function removes the 2-pion symmetrizations.

With 4-pion correlations, two types of partial cumulants as well as the full cumulant correlation function are defined:
{\footnotesize
\begin{eqnarray}
\centering
{\bf a}_4 &=& [N_4^{\rm QS}(p_1, p_2, p_3, p_4) - \epsilon_1N_2^{\rm QS}(p_1,p_2)N_1(p_3)N_1(p_4) \nonumber \\ 
&+& \epsilon_1N_1^4] / N_1^4, \label{eq:c4prime} \\
{\bf b}_4 &=& [N_4^{\rm QS}(p_1, p_2, p_3, p_4) - 3N_2^{\rm QS}(p_1,p_2)N_2^{\rm QS}(p_3,p_4) \nonumber \\
&+& 3N_1^4] / N_1^4, \label{eq:c4primeprime} \\ 
{\bf c}_4 &=& [N_4^{\rm QS}(p_1, p_2, p_3, p_4) - \epsilon_2N_3^{\rm QS}(p_1,p_2,p_3)N_1(p_4) \nonumber \\
&-& \epsilon_3N_2^{\rm QS}(p_1,p_2)N_2^{\rm QS}(p_3,p_4) + \epsilon_4N_2^{\rm QS}(p_1,p_2)N_1(p_3)N_1(p_4) \nonumber \\
&+& (\epsilon_2+\epsilon_3-\epsilon_4)N_1^4] / N_1^4. \label{eq:c4}
\end{eqnarray}}
The cumulant and partial cumulants isolate different sequences of symmetrization.
When a mixture of identical and non-identical pions is used (e.g.\ $\pi^+$ and $\pi^-$ mixtures), one need only subtract the symmetrizations from identical pion pairs and triplets.
For same-charge quadruplets the coefficients are: $\epsilon_1=6, \epsilon_2=4, \epsilon_3=3, \epsilon_4=12$.
For mixed-charge quadruplets of type 1 ($\mp\pm\pm\pm$) the coefficients are: $\epsilon_1=3, \epsilon_2=1, \epsilon_3=0, \epsilon_4=0$.
For mixed-charge quadruplets of type 2 ($\mp\mp\pm\pm$) the coefficients are: $\epsilon_1=2, \epsilon_2=0, \epsilon_3=1, \epsilon_4=0$.

In the case of same-charge quadruplets, the first partial cumulant, ${\rm {\bf a}}_4$, is defined such that all six terms of 2-pion symmetrization are removed. 
The second partial cumulant, ${\rm {\bf b}}_4$, further removes the three 2-pair symmetrization terms.
Note that $N_2^2$ contains two terms of 2-pion symmetrization as well as one term of 2-pair symmetrization.
Finally, the cumulant further removes the eight terms of 3-pion symmetrization and is thus a measure of the genuine 4-pion symmetrization alone,
Note that $N_3$ contains six terms of 2-pion symmetrization and two terms of 3-pion symmetrization.
The above equations for the cumulant and partial cumulants are valid for identical pions.  
When a mixture of identical and non-identical pions is used (e.g.\ $\pi^+$ and $\pi^-$ mixtures), one need only subtract the symmetrizations from identical pion pairs and triplets.

The cumulant and partial cumulant correlation functions have an advantage over the full correlation functions (Eq.~\ref{eq:Cn}) in low multiplicity collision events as pointed out in Ref.~\cite{Abelev:2014pja}.  
In such events, the contributions from non-Bose-Einstein correlations (e.g.\ mini-jets) have a non-negligible effect.  
Two-pion background correlations are explicitly removed with the cumulant and partial cumulants.
The remaining higher order background correlations were shown to be very small in Ref.~\cite{Abelev:2014pja}.
Also note that the choice of the $Q_3$ projection variable may also influence the flatness of the baseline.  
QS correlations are localized at low $Q_3$ while 3-pion mini-jet correlations may be less well localized with such a variable (smeared).

\subsubsection{Practicality: Nested loops}
In high multiplicity events such as those produced in heavy-ion collisions, the analysis of multi-particle correlations directly with nested loops is prohibitively expensive in terms of CPU time.
One can circumvent this problem by only retaining pion-pairs in a specific region of $q$.
A system of switches using 2D arrays of Boolean variables can be formed for each type of pion pair (e.g.\ pion-1 from event A and pion-2 from event B).
Two classes of switch-arrays should be formed: one for the femtoscopic analysis region at low q and one for the normalization region at high q.
The cutoff for the low q region should be chosen to include the dominant region of QS correlations.
The width of the normalization region should be chosen to contain sufficient statistics while still minimizing CPU time.
Nested loops are still utilized for the full analysis except that at the start of $2^{\rm nd}$ and higher loops one checks each pair with the switch-array.
If any pair is turned ``off'', one skips the entire n-tuple.
In this manner, undesired n-tuples are skipped before any time consuming calculations are performed.
The algorithm is depicted with the following pseudo-code.
\begin{algorithm}[H]
\begin{algorithmic}
    \STATE \textbf{Stage one:} Set pair Boolean switch
    \FORALL{$i=0$ to $N$}
    \FORALL{$j=i+1$ to $N$}
    \IF{$q_{\rm min} \leq q < q_{\rm max}$}
    \STATE SWITCH[$i$][$j$] = 1
    \ELSE
    \STATE SWITCH[$i$][$j$] = 0
    \ENDIF
    \ENDFOR
    \ENDFOR
    \STATE
    \STATE \textbf{Stage two:} Perform 4-pion correlation analysis
    \FORALL{$i=0$ to $N$}
    \FORALL{$j=i+1$ to $N$}
    \IF{SWITCH[$i$][$j$] == 0} \STATE skip
    \ENDIF
    \FORALL{$k=j+1$ to $N$}
    \IF{SWITCH[$i$][$k$] == 0} \STATE skip
    \ENDIF
    \IF{SWITCH[$j$][$k$] == 0} \STATE skip
    \ENDIF
    \FORALL{$l=k+1$ to $N$}
    \IF{SWITCH[$i$][$l$] == 0} \STATE skip
    \ENDIF
    \IF{SWITCH[$j$][$l$] == 0} \STATE skip
    \ENDIF
    \IF{SWITCH[$k$][$l$] == 0} \STATE skip
    \ENDIF
    \STATE \textbf{Store quadruplet}
    \ENDFOR
    \ENDFOR
    \ENDFOR
    \ENDFOR
\end{algorithmic}
\end{algorithm}

\section{THERMINATOR calculations} \label{sec:THERM}
In this section several calculations of 3- and 4-pion correlations in the heavy-ion generator \textsc{therminator} 2 \cite{Kisiel:2005hn,Chojnacki:2011hb} are provided.
\textsc{therminator} 2 is a Monte Carlo event generator based on the statistical hadronization of partons at kinetic freeze-out.
A small number of physical input parameters are used such as the temperature, chemical potentials, initial size, and the velocity of collective flow which has been shown to describe a great deal of the RHIC and LHC observables.
An important feature of \textsc{therminator} is the inclusion of the full set of hadronic resonances.

A 3+1D viscous hydrodynamic hypersurface is input into the statistical hadronization process.
Events were generated with the following settings: Pb--Pb collisions at $\sqrt{s_{\rm NN}} = 2.76$ TeV, $b=2.3$ fm (impact parameter), $T_{i}=512$ MeV (initial central temperature), $t_0=0.6$ fm (starting time of hydrodynamics), $T_{f}=140$ MeV (freeze-out temperature).
Approximately $270\times10^3$ events were generated.
Only charged pions were retained with the following criteria: $0.16 < p_{\rm T} < 1.0$ GeV/$c$, $|\eta|<0.8$.
To simulate the typical experimental resolution of a track's distance-of-closest approach (DCA) to the primary vertex, pions which freeze-out greater than 1 cm from the collision center are rejected.
For the correlation part of this analysis, only the dominant region of QS correlations are analyzed by requiring a cut on the pair relative separation in the PRF: $0.1<r^*<100$ fm.

\subsection{$f$ coefficients}
\begin{figure}
\center
  \includegraphics[width=0.49\textwidth]{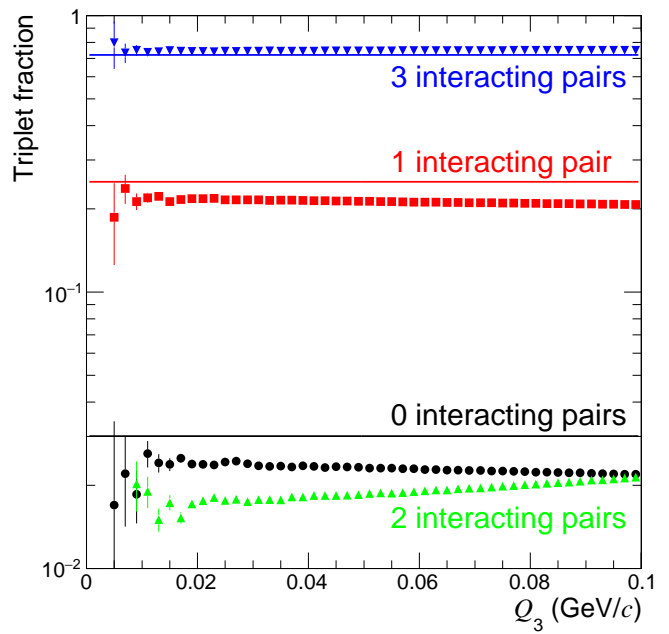}
  \caption{(Color online) Triplet fractions in \textsc{therminator}.  Blue triangles (3 interacting pairs), red squares (1 pair), green triangles (2 pairs), and black circles (0 pairs) are shown.  The core/halo values (given $f_c^2=0.803$) are shown with blue (3 pairs), red (1 pair), and black (0 pairs) lines.}
  \label{fig:TripletFractions}
\end{figure}

\begin{figure}
\center
  \includegraphics[width=0.49\textwidth]{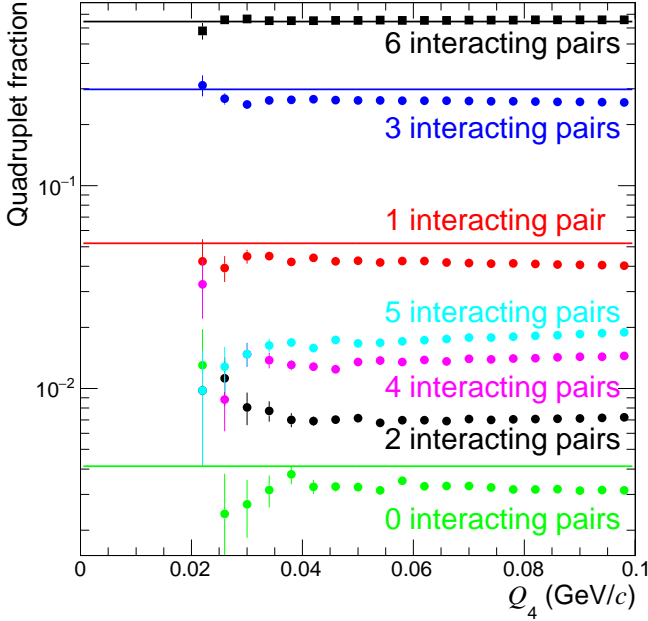}
  \caption{(Color online) Quadruplet fractions in \textsc{therminator}.  From top to bottom, the markers indicate 6, 3, 1, 5, 4, 2, and 0 interacting pair(s).  The core/halo values (given $f_c^2=0.803$) are shown with black (6 interacting pairs), blue (3 pairs), red (1 pair), and green (0 pairs) lines.}
  \label{fig:QuadFractions}
\end{figure}

The $f$ coefficients in Eqs.~\ref{eq:N3} and \ref{eq:N4} characterize the probabilities of short- and long-range interactions for triplets and quadruplets. 
A pair is deemed interacting in \textsc{therminator} if its value of $r^*$ is less than a certain cutoff, which was taken to be 80 fm (well above the dominant QS region).
With this cutoff, the fraction of interacting pairs is $f_c^2=0.803$.
The triplet fractions are shown in Fig.~\ref{fig:TripletFractions}.
The blue, red, and black lines correspond to the core/halo values of $f_{33}$, $3f_{32}$, and $f_{31}$, respectively.
Note that the case of 2 interacting pairs is absent in the core/halo picture but non-zero in general.
One example of such a case would be a linear configuration of 3 pions where the middle pion is separated from both end pions by an amount less than the cutoff while the end pions themselves are separated by a greater distance.
The quadruplet fractions are shown in Fig.~\ref{fig:QuadFractions}.
The black, blue, red, and green lines correspond to the core/halo values of $f_{44}$, $4f_{43}$, $6f_{42}$, and $f_{41}$ respectively.
Note that the cases of 5, 4, and 2 interacting pairs are absent in the core/halo picture but non-zero in general.
The case of 5 interacting pairs can occur for a parallelogram configuration of the 4 pions where the longest axis represents the length above the $r^*$ cut. 
An example of a case with 4 interacting pairs can occur with an equilateral triangle of 3 pions and the fourth pion being located outside of the triangle but near one of the vertices.  
The case of 2 interacting pions can a be a linear configuration of 3 pions and the fourth being far away from the rest.
For both triplet and quadruplet fractions there is no significant $Q_3$ and $Q_4$ dependence.  
Thus, the $f$ coefficients are largely independent of relative momentum.

In experiment, one cannot isolate each of the possibilities with mixed-event techniques.
For instance, the case of 2 interacting pairs in the triplet cannot be isolated using the three available distributions: $N_3(p_1,p_2,p_3)$, $N_2(p_1,p_2)N_1(p_3)$, and $N_1(p_1)N_1(p_2)N_1(p_3)$.
The modification to the core/halo $f$ coefficients in Eqs.~\ref{eq:N3} and \ref{eq:N4} are estimated by combining the appropriate \textsc{therminator} fraction types.  
For triplets, the modification to $f_{33}$ is given by the sum of 2 and 3 interacting pair fractions in Fig.~\ref{fig:TripletFractions}.
For quadruplets, the modification to $f_{44}$ is given by the sum of 4, 5, and 6 interacting pair fractions in Fig.~\ref{fig:QuadFractions}.
The modification to $f_{43}$ is given by the sum of 2 and 3 interacting pair fractions.
Table \ref{tab:TripletMods} shows the core/halo values of $f_{33},f_{32},f_{31}$ and their percentage modification in \textsc{therminator} in parenthesis.
The calculation is done for three different choices of the $r^*$ cutoff.
\begin{table}
  \center
{\footnotesize 
  \begin{tabular}{| c | c | c | c |}
    \hline
    Triplets & $f_{33}$ & $f_{32}$ & $f_{31}$ \\ \hline
    $r^*<60$ & 0.675($+9\%$) & 0.094($-17\%$) & 0.041($-25\%$)   \\ \hline
    $r^*<80$ & 0.720($+6\%$) & 0.083($-15\%$) & 0.030($-24\%$)   \\ \hline
    $r^*<100$ & 0.738($+6\%$) & 0.079($-15\%$) & 0.026($-24\%$)   \\ \hline
  \end{tabular}
}
  \caption{Triplet $f$ factors in the core/halo picture and the \textsc{therminator} modification percentage in parenthesis.  Note that $3f_{32}$ is shown in Fig.~\ref{fig:TripletFractions}.}
  \label{tab:TripletMods}
\end{table}
Table \ref{tab:QuadMods} shows the core/halo values of $f_{44},f_{43},f_{42},f_{41}$ and their percentage modifications in \textsc{therminator} in parenthesis.
\begin{table}
  \center
{\footnotesize
  \begin{tabular}{| c | c | c | c | c |}
    \hline
    Quadruplets & $f_{44}$ & $f_{43}$ & $f_{42}$ & $f_{41}$ \\ \hline
    $r^*<60$ & 0.593($+10\%$) & 0.083($-12\%$) & 0.012($-26\%$) & 0.007($-18\%$)  \\ \hline
    $r^*<80$ & 0.645($+6\%$) & 0.075($-9\%$) & 0.009($-19\%$) & 0.004($-11\%$)  \\ \hline
    $r^*<100$ & 0.667($+5\%$) & 0.071($-8\%$) & 0.008($-18\%$) & 0.003($-7\%$)  \\ \hline
  \end{tabular}
}
  \caption{Quadruplet $f$ factors in the core/halo picture and the \textsc{therminator} modification percentage in parenthesis.  Note that $4f_{43}$ and $6f_{42}$ are shown in Fig.~\ref{fig:QuadFractions}.}
  \label{tab:QuadMods}
\end{table}

\subsection{Measured and built correlation functions}
\begin{figure}
\center
  \includegraphics[width=0.49\textwidth]{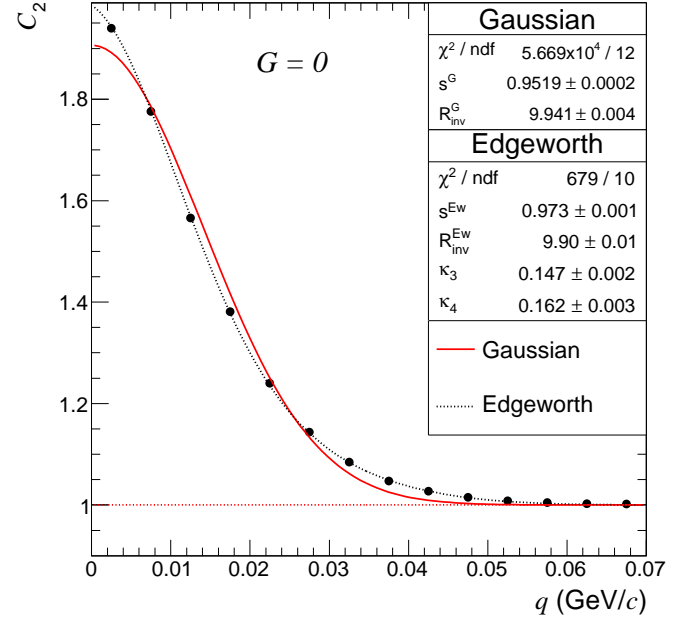}
  \caption{(Color online) $C_2$ versus $q$ calculated in \textsc{therminator}.  A Gaussian as well as Edgeworth fit is shown.  The fit range is that which is shown in the figure.  $0.2<\kT<0.3$ GeV/$c$.}
  \label{fig:C2therm}
\end{figure}

The 2-, 3-, and 4-pion QS correlation functions in \textsc{therminator} are now presented.
The correlation functions are calculated via the fully symmetrized 2-, 3-, and 4-pion plane-wave functions.
The symmetrized 2-pion plane-wave function is given by $\Psi_2 = \frac{1}{\sqrt{2}}[e^{-i{\bf k^*r^*}} + e^{i{\bf k^*r^*}}]$.
The symmetrized 3-pion plane-wave function was given in Eq.~\ref{eq:Psi3} with $\Phi=1$.
In appendix A the symmetrized 4-pion plane-wave function is given.
The modulus square of the plane-wave functions is applied as a pair/triplet/quadruplet weight and is averaged over the freeze-out space-time coordinates in each event.
The weight is then averaged over all events.
The 2-pion correlation function is shown in Fig.~\ref{fig:C2therm}.
It is fit with a Gaussian and Edgeworth parametrization: $C_2(q) = 1 + s^2 E_{\rm w}^2(R\,q)\,e^{-R^2q^2}$. 
The Gaussian case corresponds to $E_{\rm w}=1$.
Note, that only femtoscopically separated particles are retained ($r^*<100$ fm) and FSI were not included in the weighting procedure.
In our calculation, due to the full symmetrization, pion production is fully chaotic and thus the $s$ parameter of the fit is just a scale factor unrelated to coherence.

\begin{figure}
\center
  \includegraphics[width=0.49\textwidth]{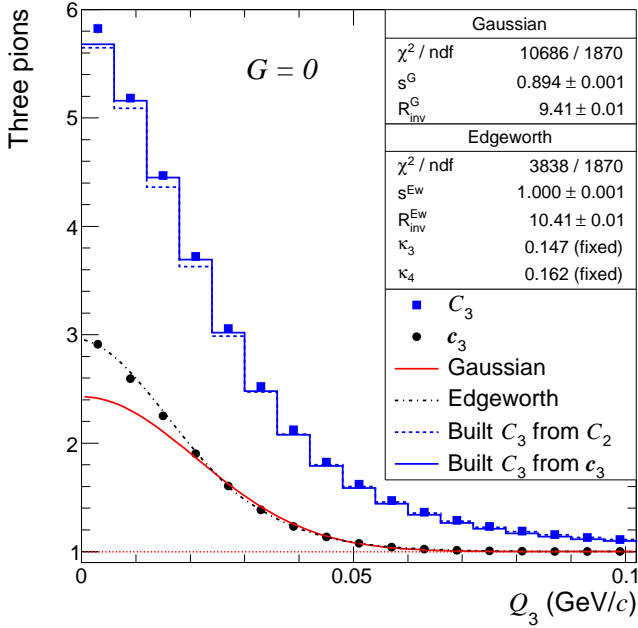}
  \caption{(Color online) Three-pion correlations versus $Q_3$ calculated in \textsc{therminator}.  A Gaussian as well as Edgeworth fit to the cumulant correlation function is shown.  The fits are performed in 3D ($q_{12},q_{23},q_{31}$).  The built correlation functions are also shown as block histograms. $0.16<\KTThree<0.3$ GeV/$c$.}
  \label{fig:C3therm}
\end{figure}

\begin{figure}
\center
  \includegraphics[width=0.49\textwidth]{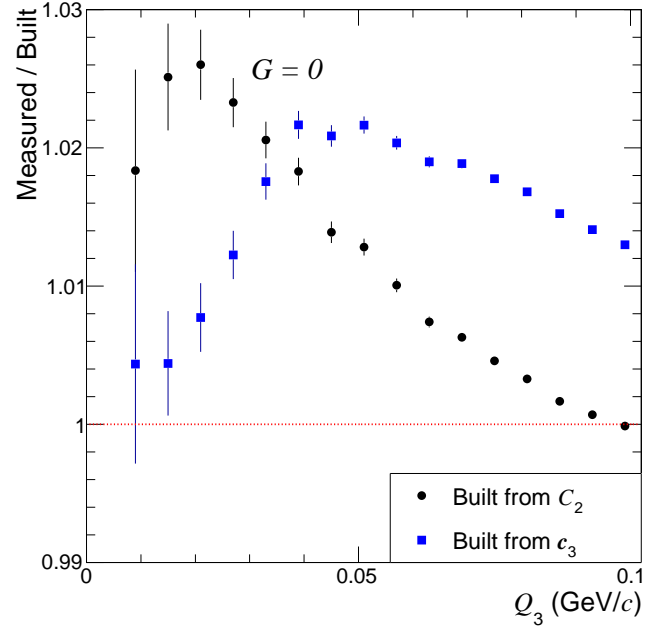}
  \caption{(Color online) Ratio of measured to built 3-pion correlation functions.}
  \label{fig:C3thermRatio}
\end{figure}

The 3-pion correlation function projected against $Q_3$ is shown in Fig.~\ref{fig:C3therm}.
A Gaussian and Edgeworth fit to the cumulant correlation is performed in 3D according to 
{\footnotesize
\begin{equation}
{\bf c}_3(q_{12},q_{23},q_{31}) = 1 + s^3 e^{-R^2(q_{12}^2+q_{23}^2+q_{31}^2)/2}E_{\rm w,12}E_{\rm w,23}E_{\rm w,31},
\end{equation}}
which can be obtained by substituting Edgeworth exchange amplitudes from Eq.~\ref{eq:EWEA} into Eq.~\ref{eq:c3EA} (neglecting the 3-pion phase).
Two types of built correlation functions are shown in Fig.~\ref{fig:C3therm}.
Dashed blue lines correspond to $C_3$ built from $T_{ij}$ extracted directly from $C_2(\kT,\qo,\qs,\ql)$.
Solid blue lines correspond to $C_3$ built from $T_{ij}$ extracted from ${\bf c}_3(q_{12},q_{23},q_{31})$ fits.
The building of correlation functions is done according to Eq.~\ref{eq:C3NoPhase} with $G=0$.
The ratio of measured to built $C_3$ is shown in Fig.~\ref{fig:C3thermRatio}.
Built correlations are found to reproduce the measured ones to within a few percent while those from $C_2(\kT,\qo,\qs,\ql)$ are found to be somewhat more accurate.
This is partly expected since 2-pion correlations are binned much more differentially.

\begin{figure}
\center
  \includegraphics[width=0.49\textwidth]{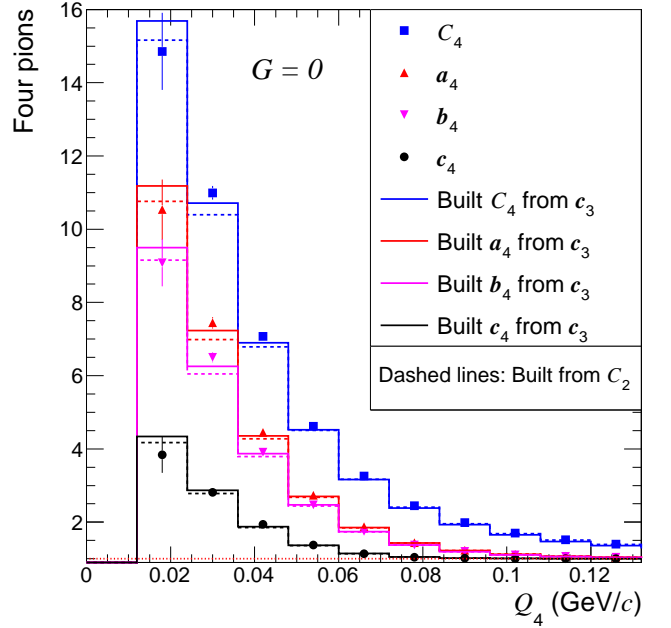}
  \caption{(Color online) Four-pion correlations versus $Q_4$ calculated in \textsc{therminator}.  The built correlation functions are also shown as a block histogram. $0.16<\KTFour<0.3$ GeV/$c$.}
  \label{fig:C4therm}
\end{figure}

\begin{figure}
\center
  \includegraphics[width=0.49\textwidth]{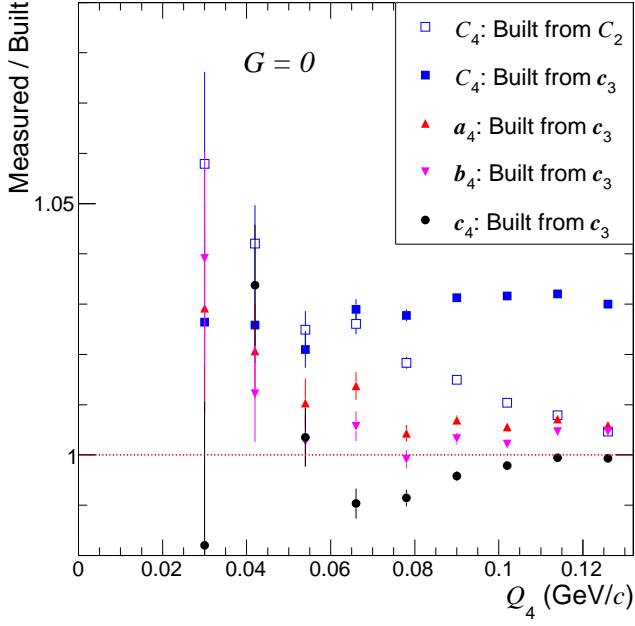}
  \caption{(Color online) Ratio of measured to built 4-pion correlation functions.}
  \label{fig:C4thermRatio}
\end{figure}

The 4-pion correlation function projected against $Q_4$ is shown in Fig.~\ref{fig:C4therm}.
The partial cumulants, ${\bf a}_4$ and ${\bf b}_4$ represent the removal of 2-pion and 2-pion+2-pair symmetrizations, respectively.
The full cumulant, ${\bf c}_4$, represents the isolation of genuine 4-pion symmetrizations.
Built correlation functions are shown with the colored box histograms.
The building of correlation functions is done according to Eq.~\ref{eq:C4NoPhase} with $G=0$.
The ratio of measured to built $C_4$ is shown in Fig.~\ref{fig:C4thermRatio}.
Similar to built 3-pion correlations, built 4-pion correlations generally under predict the measured ones by a few percent.
The reason for the slight bias is investigated further in Sec.~\ref{sec:r3} with $r_3$ and is related to the finite binning of multi-dimensional correlation functions.
For the built correlations in this subsection, 2-pion correlations are binned as follows: 4 \kT~bins dividing the full \kT~range, 5 MeV$/c$ bin widths for \qo,\qs,\ql.  
Linear interpolation is used in the \kT~dimension while cubic is used in the other dimensions. 
Three-pion cumulants are binned with 5 MeV/$c$ widths in each pair $q_{ij}$.

\subsubsection{Including partial coherence} \label{sec:IncludeG}
To demonstrate the building of QS correlations with partial coherence, a coherent source of pions is inserted in \textsc{therminator}.
Pions are randomly reassigned such that $35\%$ of the pions in an event originate from a spherical Gaussian coherent source with $R_{\rm coh}=1$ fm.
That is, $G=0.35$ and $t_{ij}=e^{-R_{\rm coh}^2q_{ij}^2/2}$.
The momentum distribution of the coherent pions is identical to that of the chaotic pions.
The expected suppression from coherence is incorporated by setting the appropriate plane-wave contributions to zero.
In the fully symmetrized plane-wave function, individual plane-waves corresponding to an $n$-pion symmetrization are set to zero if more than $n-1$ of the pions are from the coherent source.

\begin{figure}
\center
  \includegraphics[width=0.49\textwidth]{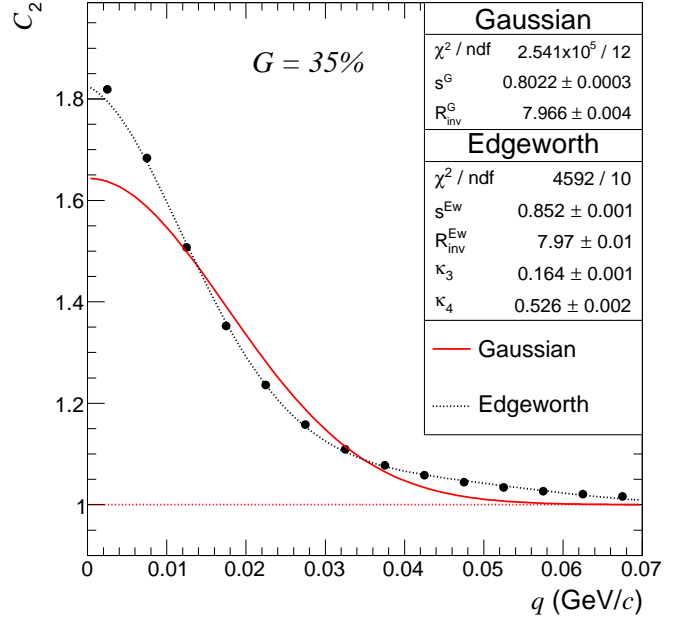}
  \caption{(Color online) $C_2$ versus $q$ calculated in \textsc{therminator} with $G=0.35$ and $t_{ij}=e^{-R_{\rm coh}^2q_{ij}^2/2}$.  A Gaussian as well as Edgeworth fit is shown.  Further details are the same as in Fig.~\ref{fig:C2therm}.}
  \label{fig:C2thermG}
\end{figure}

\begin{figure}
\center
  \includegraphics[width=0.49\textwidth]{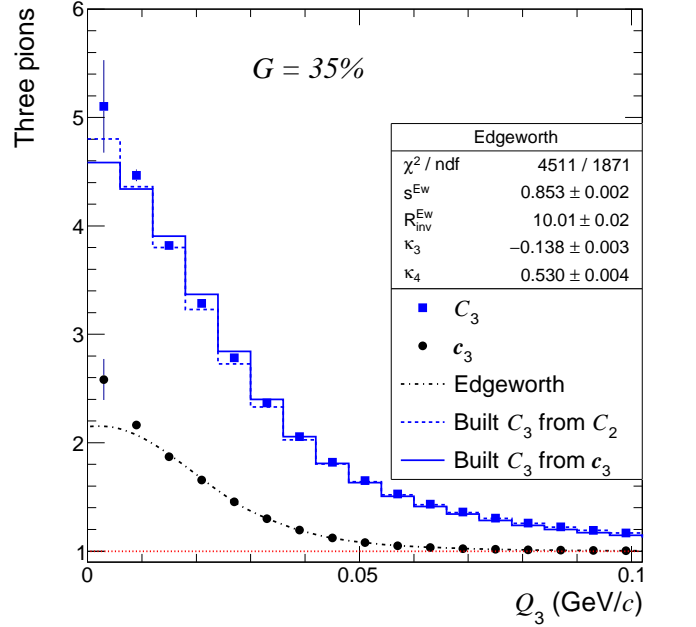}
  \caption{(Color online) Three-pion correlations versus $Q_3$ calculated in \textsc{therminator} with $G=0.35$ and $t_{ij}=e^{-R_{\rm coh}^2q_{ij}^2/2}$.  An Edgeworth fit is shown.  Further details are the same as in Fig.~\ref{fig:C3therm}.}
  \label{fig:C3thermG}
\end{figure}

\begin{figure}
\center
  \includegraphics[width=0.49\textwidth]{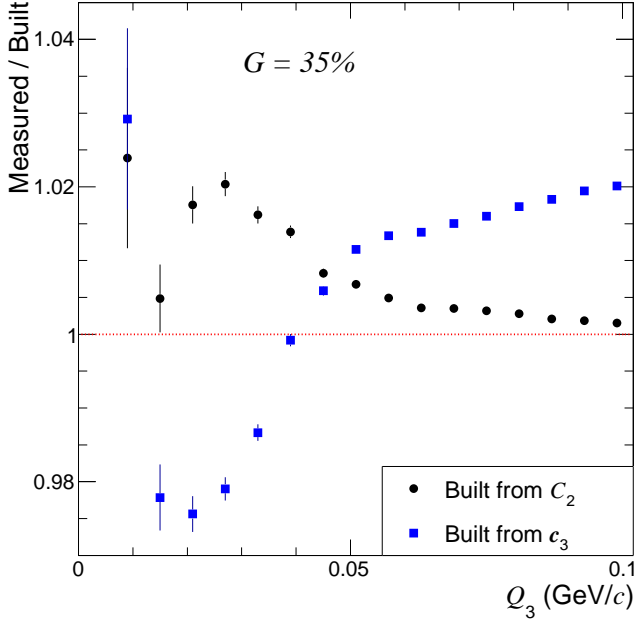}
  \caption{(Color online) Ratio of measured to built 3-pion correlation functions with $G=0.35$ and $t_{ij}=e^{-R_{\rm coh}^2q_{ij}^2/2}$.}
  \label{fig:C3thermRatioG}
\end{figure}

\begin{figure}
\center
  \includegraphics[width=0.49\textwidth]{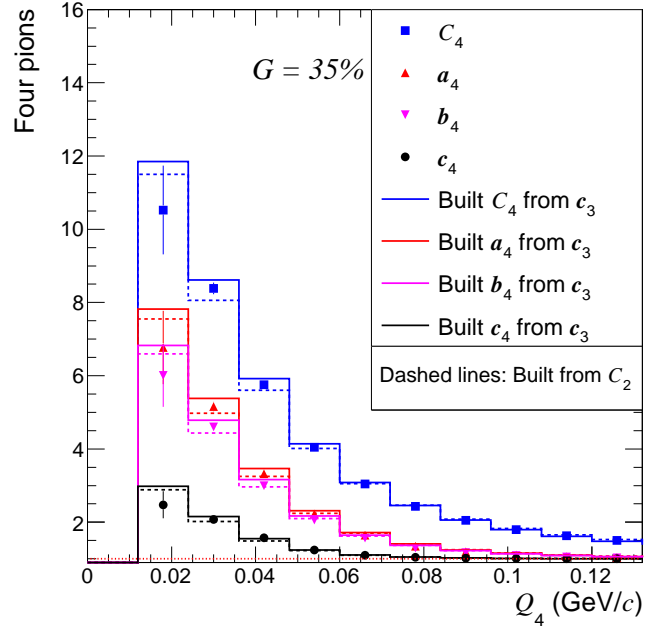}
  \caption{(Color online) Four-pion correlations versus $Q_4$ calculated in \textsc{therminator} with $G=0.35$ and $t_{ij}=e^{-R_{\rm coh}^2q_{ij}^2/2}$.  Further details are the same as in Fig.~\ref{fig:C4therm}.}
  \label{fig:C4thermG}
\end{figure}

\begin{figure}
\center
  \includegraphics[width=0.49\textwidth]{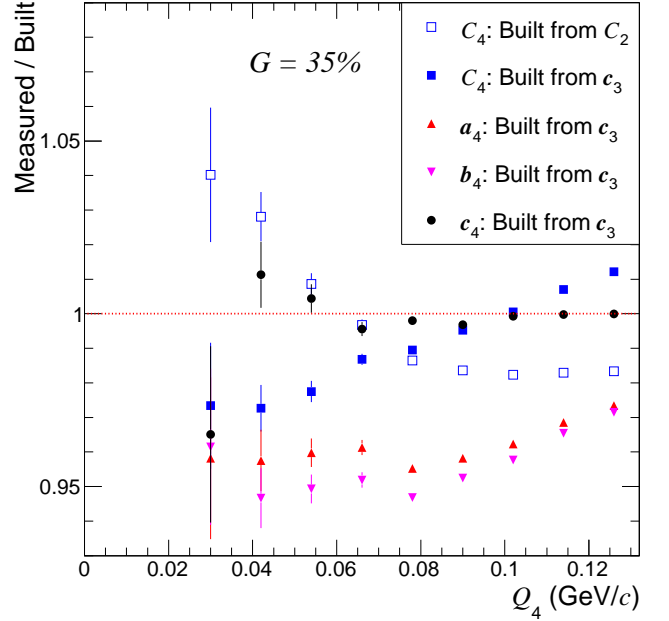}
  \caption{(Color online) Ratio of measured to built 4-pion correlation functions with $G=0.35$ and $t_{ij}=e^{-R_{\rm coh}^2q_{ij}^2/2}$.}
  \label{fig:C4thermRatioG}
\end{figure}

The 2-pion correlation function with the parametrized coherent source is shown in Fig.~\ref{fig:C2thermG}.
The 3-pion correlation function with the parametrized coherent source is shown in Fig.~\ref{fig:C3thermG}.
An Edgeworth fit to the cumulant correlation is shown with a dashed blue line.
The fit well describes the cumulant for $Q_3>0.015$ GeV/$c$.
Below that value, the fit is below the data and is part of the reason why the built correlation from ${\bf c}_3$ underestimates the measured $C_3$ in the same region.
The built correlation function from $C_2$ is also beneath the measured $C_3$ for reasons described in Sec.~\ref{sec:r3}.
The ratio of measured to built $C_3$ is shown in Fig.~\ref{fig:C3thermRatioG}.
The 4-pion correlation function with the parametrized coherent source is shown in Fig.~\ref{fig:C4thermG}.
The ratio of measured to built $C_4$ is shown in Fig.~\ref{fig:C4thermRatioG}.
The built correlations incorporate the known $G$ and $t_{ij}$ values and thus illustrate how built correlations can be used to estimate the coherent fraction.

\subsection{$r_3$ and $r_4$ measurements} \label{sec:r3}
The 3- and 4-pion phase factors in \textsc{therminator} are measured with $r_3$ and $r_4$ in Eqs.~\ref{eq:r3} and \ref{eq:r4}.
In the computation of both denominators, the pair exchange magnitudes appear.  
As this quantity must be evaluated numerically by averaging over the sources produced in many events, it is tabulated discretely in the first pass over the data.
In the second pass one interpolates between the bins linearly or cubically. 
Finite bin width is found to cause a bias on $r_3$ and $r_4$ in the same manner as it occurred for the built correlation functions.
The bias is largest for linear interpolation between widely spaced bins of $q_{\rm out},q_{\rm side}$, and $q_{\rm long}$ and depends on the concavity of the correlation function.
For instance, negative concavity with linear interpolation generally leads to an underestimation.
The concavity of a Gaussian correlation function is negative for $q<\sqrt{1/2R^2}$ and positive for larger $q$.
The bias may be reduced somewhat with a cubic interpolator.

\begin{figure}
\center
  \includegraphics[width=0.49\textwidth]{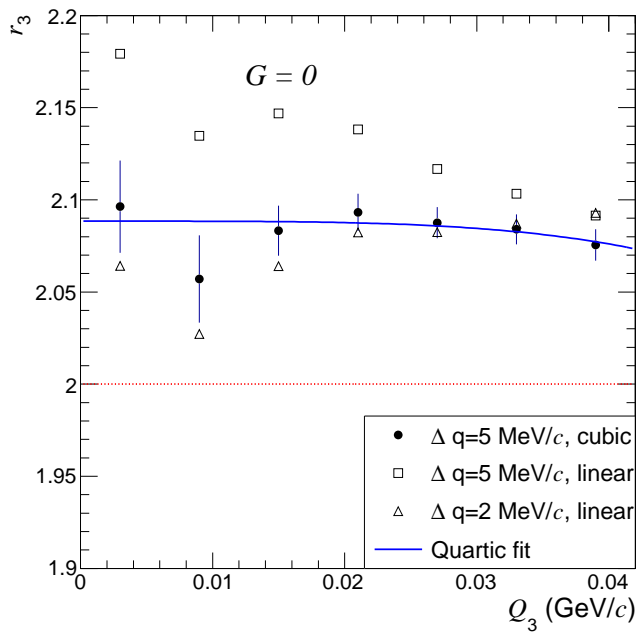}
  \caption{(Color online) $r_3$ calculated in \textsc{therminator} with various types of $T_{ij}$ tabulations/interpolations.  The quartic fit is applied to the solid points.  Statistical errors are only drawn for the solid points and are the same for the others.  The chaotic upper limit is shown with the red dotted line.}
  \label{fig:r3}
\end{figure}

\begin{figure}
\center
  \includegraphics[width=0.49\textwidth]{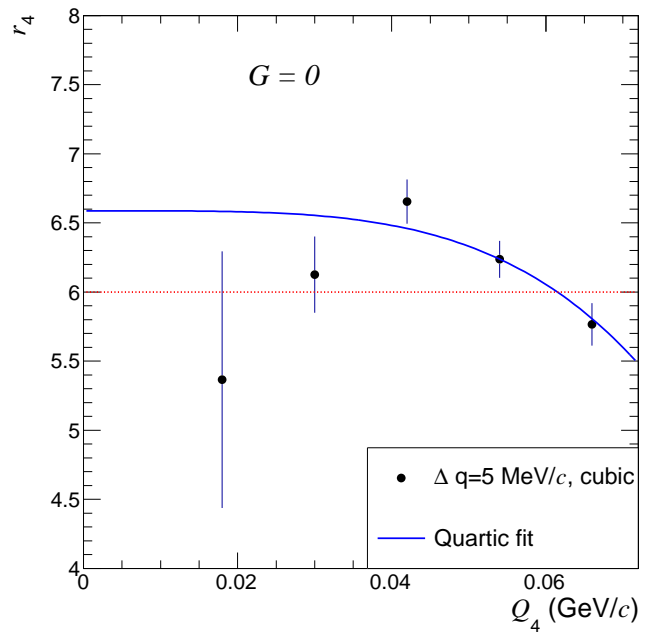}
  \caption{(Color online) $r_4$ calculated in \textsc{therminator} with various types of $T_{ij}$ tabulations/interpolations.  The chaotic upper limit is shown with the red dotted line.}
  \label{fig:r4}
\end{figure}

The 2-pion $T_{ij}$ factors are computed in 3 different ways depending on the number of $q$ bins and on the interpolation scheme between $q$ bins (linear or cubic).  
The $\kT$ dependence of $T_{ij}$ was found to be fairly linear and thus linear interpolation is always used between $\kT$ bins.
Three different $q$ bin widths (1, 2, or 5 MeV/$c$) are tried while dividing the full $\kT$ interval into 4 bins.
In Fig.~\ref{fig:r3} and \ref{fig:r4}, $r_3$ and $r_4$ calculations in \textsc{therminator} are shown.
Similar to Ref.~\cite{Abelev:2013pqa}, quartic ($r_n(Q_n)=I(1-aQ_n^4)$) fits are shown for both $r_3$ and $r_4$ for $\Delta q=5$ MeV/$c$, cubic.
The quartic intercept parameter to $r_3$ is $I=2.08\pm0.01$ while $a$ is consistent with zero.
For $r_4$ they are $I=6.6\pm0.1,a=(6\pm2)\times10^3$.
The near independence of $r_3$ and $r_4$ with $Q_3$ and $Q_4$ indicate a negligible effect of 3- and 4-pion phases in \textsc{therminator}.
This feature is expected for the case where the space-time point of maximum pion emission is momentum independent \cite{Heinz:1997mr}.
In Fig.~\ref{fig:r3} one also observes that the chaotic upper limit (2.0) is exceeded to varying degrees.  
The bias is caused by the limited dimensionality and finite bin widths of $T_{ij}$, especially in the case of linear interpolation.
Cubic interpolation is observed to reduce the bias.

\begin{figure}
\center
  \includegraphics[width=0.49\textwidth]{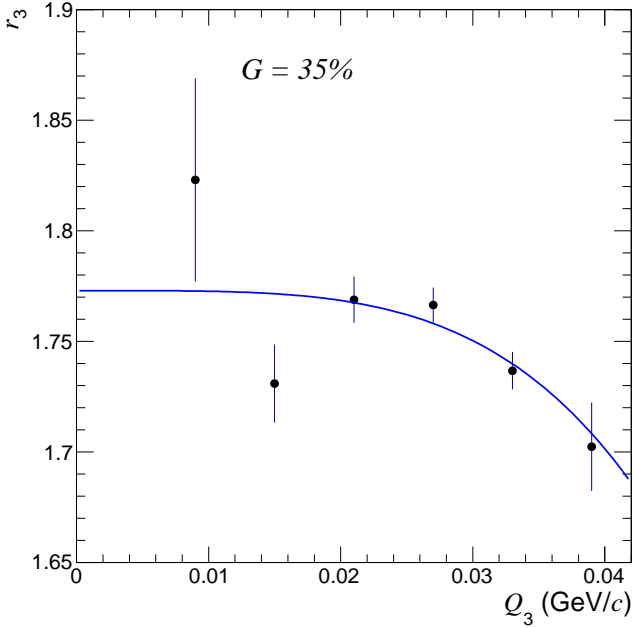}
  \caption{(Color online) $r_3$ calculated in \textsc{therminator} with $G=0.35$ and $t_{ij}=e^{-R_{\rm coh}^2q_{ij}^2/2}$.  Bin widths are 5 MeV/$c$ with cubic interpolation.  A quartic fit is shown. Statistical errors are only drawn for the solid points and are the same for the others.}
  \label{fig:r3G}
\end{figure}

With a coherent component inserted as in Sec.~\ref{sec:IncludeG}, the expected suppression of $r_3$ is observed as seen in Fig.~\ref{fig:r3G}.
The quartic fit parameters in Fig.~\ref{fig:r3G} are: $I=1.77\pm0.01$ and $a=(1.6\pm0.5)\times10^4$.
The intercept parameter is related to the coherent fraction as
\begin{equation}
I = 2\sqrt{1-G}\frac{1+2G}{(1+G)^{3/2}},
\end{equation}
which yields $G=0.33\pm0.01$.
Note that in the recent ALICE $r_3$ measurement \cite{Abelev:2013pqa}, linear interpolation was used with 4 $\kT$ bins and 5 MeV/$c$ $q$ bin widths.
Correction for the finite binning effect may lower $r_3$ and increase the extracted coherent fraction.
Concerning the $a$ parameter, it was shown in Ref.~\cite{Heinz:1997mr} that quartic behavior is expected when the space-time point of maximum pion emission is momentum dependent.
As the $a$ parameter for the chaotic part of the \textsc{therminator} source is zero in Fig.~\ref{fig:r3}, and since the momentum spectrum of the coherent and chaotic components are identical, one expects $a$ to vanish also with the inserted coherent component.
The non-vanishing value of $a$ also characterizes the same bias as attributed to $I$.

\subsection{3-pion FSI calculations}
\begin{figure}
\center
  \includegraphics[width=0.49\textwidth]{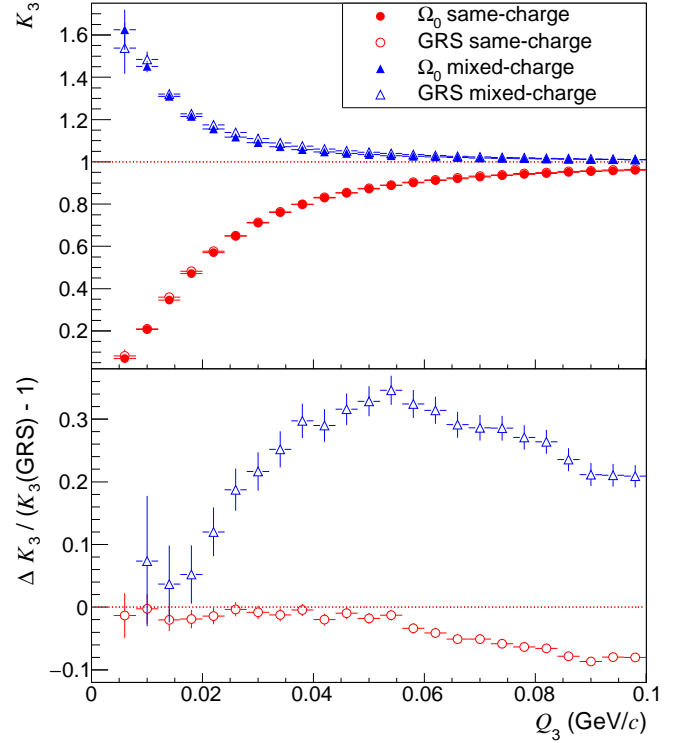}
  \caption{(Color online) Comparison of same and mixed-charge three-pion FSI correlations.  $\Omega_0$ and generalized Riverside (GRS) methods are shown. 
    The calculation was performed in \textsc{therminator}.  The bottom panel
    shows the difference between the two methods, $\Delta K_{3}=K_{3}(\Omega_{0})-K_{3}(\rm GRS)$, divided by $K_{3}(\rm GRS)-1$.}
  \label{fig:K3comp}
\end{figure}

In Ref.~\cite{Abelev:2013pqa} and Fig.~\ref{fig:K3comp}, a comparison between $\Omega_0$ and GRS 3-pion FSI correlation is made.  
One observes a striking similarity between the two methods, especially for same-charge triplets.  
Less similar is the mixed-charge calculation for which the {\it cumulants} differ by as much as $30\%$.
In Ref.~\cite{Abelev:2013pqa}, the mixed-charge calculations were more similar and is due to the inclusion of strong FSI in the factorization ansatz.
The calculation in Fig.~\ref{fig:K3comp} excludes strong FSI.
The similarity of the GRS and $\Omega_0$ method is related to the factorizability of separate $F$ factors as well as that between $F$ and plane-wave factors.
As one decreases the source size in the calculation, both methods converge as the $F$ factors approach unity.

\section{Multi-boson distortions}
The standard framework of Bose-Einstein interferometry neglects the effect of higher-order symmetrizations and is only valid in the limit of low phase-space densities at freeze-out \cite{Zajc:1986sq,Pratt:1993uy,Pratt:1994cg,Lednicky:1999xz,Zhang:1998sz}.  
In such a dilute pion gas, $n$-pion correlation functions are calculable purely in terms of $n$-pion symmetrizations.
At higher densities, the higher-order symmetrizations become significant and {\it distort} the correlation functions.
Previous calculations revealed a widened 2-pion correlation function as well as a suppressed intercept, both of which are unrelated to quantum coherence.
Here, the effect of the distortions on the comparison of built and measured 3- and 4-pion correlation functions is analyzed.

The computation of all higher orders of symmetrizations is greatly simplified when the $n$-pion emission function is assumed to factorize into a product of single particle emission functions $S(x,p)$.
A remaining complication is the computation of the $n!$ plane-wave functions which can be alleviated with the help of the Pratt ring integrals \cite{Pratt:1993uy,Pratt:1994cg}
{\footnotesize 
\begin{eqnarray}
G_1({\bf p}_1,{\bf p}_2) &=& \int d^4x S(\frac{1}{2}(p_1+p_2), x) e^{i(p_1-p_2)x}, \\
G_n({\bf p}_1,{\bf p}_2) &=&\int d^3{\bf k}_2 ... d^3{\bf k}_n G_1({\bf p}_1,{\bf k}_2)...G_1({\bf k}_n,{\bf p}_2). 
\end{eqnarray}}
With a Gaussian ansatz for the single particle emission function, the ring integrals are given analytically in Ref.~\cite{Lednicky:1999xz} and are used here to estimate the distortion to 3- and 4-pion correlation functions.
The single and 2-pion spectrum at a fixed multiplicity $n$ are given in Ref.~\cite{Lednicky:1999xz}
{\footnotesize 
\begin{eqnarray}
N_n^{(1)}({\bf p}) &=& \sum_{j=0}^{n-1}\frac{ \omega_{n-1-j}/(n-1-j)! }{ \omega_n/n!} G_{(j+1)}({\bf p}, {\bf p}) \\
N_n^{(2)}({\bf p}_1, {\bf p}_2) &=& \sum_{j=0}^{n-2}\frac{ \omega_{n-2-j}/(n-2-j)! }{ \omega_n/n!} \nonumber \\
&\times& \sum_{l=0}^{j} G_{l+1}({\bf p}_1, {\bf p}_1) G_{j-l+1}({\bf p}_2, {\bf p}_2) \nonumber \\
&+& G_{l+1}({\bf p}_1, {\bf p}_2) G_{j-l+1}({\bf p}_2, {\bf p}_1)
\end{eqnarray}}
where $\omega_n$ is the Bose-Einstein weight of an event with $n$ identical bosons.
Extending the techniques in Ref.~\cite{Lednicky:1999xz}, the order $i$ spectra are obtained by the appropriate symmetrization of $i$ $G$ factors.
The 3-pion spectrum is given by
{\footnotesize
\begin{eqnarray}
&&N_n^{(3)}({\bf p}_1, {\bf p}_2, {\bf p}_3) = \sum_{j=0}^{n-3}\frac{ \omega_{n-3-j}/(n-3-j)! }{ \omega_n/n!} \sum_{l=0}^{j} \sum_{m=0}^{j-l} \nonumber \\
&&\sum_{\alpha^{(3)}}G_{l+1}({\bf p}_1, {\bf p}_{\alpha_1}) G_{m+1}({\bf p}_2, {\bf p}_{\alpha_2}) G_{j-l-m+1}({\bf p}_3, {\bf p}_{\alpha_3}), 
\end{eqnarray}}
where the set of all permutations of the 3 pions is given by $\alpha^{(3)}$ and $\alpha_{1,2,3}$ represent the permuted indices.
The 4-pion spectrum is given by
{\footnotesize
\begin{eqnarray}
&&N_n^{(4)}({\bf p}_1, {\bf p}_2, {\bf p}_3, {\bf p}_4) = \sum_{j=0}^{n-4}\frac{ \omega_{n-4-j}/(n-4-j)! }{ \omega_n/n!} \sum_{l=0}^{j} \sum_{m=0}^{j-l} \sum_{s=0}^{j-l-m}  \nonumber \\
&&\sum_{\alpha^{(4)}} G_{l+1}({\bf p}_1, {\bf p}_{\alpha_1}) G_{m+1}({\bf p}_2, {\bf p}_{\alpha_2}) \nonumber \\
&&\times G_{s+1}({\bf p}_3, {\bf p}_{\alpha_3}) G_{j-l-m-s+1}({\bf p}_4, {\bf p}_{\alpha_4}), 
\end{eqnarray}}
where the set of all permutations of the 4 pions is given by $\alpha^{(4)}$ and $\alpha_{1,2,3,4}$ represent the permuted indices.
The normalization constant for the correlation functions are obtained by the term with the lowest order ring integrals ($G_1$).
The 3- and 4-pion normalizations are
\begin{eqnarray}
\frac{n^2 \omega_{n-1}^3}{(n-1)(n-2) \omega_n^2 \omega_{n-3}} \\
\frac{n^3 \omega_{n-1}^4}{(n-1)(n-2)(n-3) \omega_n^3 \omega_{n-4}} 
\end{eqnarray}

In the Gaussian model of Ref.~\cite{Lednicky:1999xz}, three parameters characterize the pion spectra: $\Delta$, $r_0$, and $n$.  
$\Delta$ characterizes the width of the momentum spectrum unaffected by the Bose-Einstein enhancement.
The isotropic spatial width of the emission function in the PRF is given by $r_0$.
To quantify the multi-boson distortions in the recent LHC central $0-5\%$ Pb--Pb data \cite{Abelev:2013pqa,Abelev:2014pja}, $\Delta=0.25$ GeV/$c$, $r_0=8.5$ fm, $n=700,1400$, and $\left<p\right>=0.2$ GeV/$c$.
The choice of $r_0$ is based on the extracted radii from 1D Gaussian fits to the measured correlation functions \cite{Abelev:2014pja}.

\begin{figure}
\center
  \includegraphics[width=0.49\textwidth]{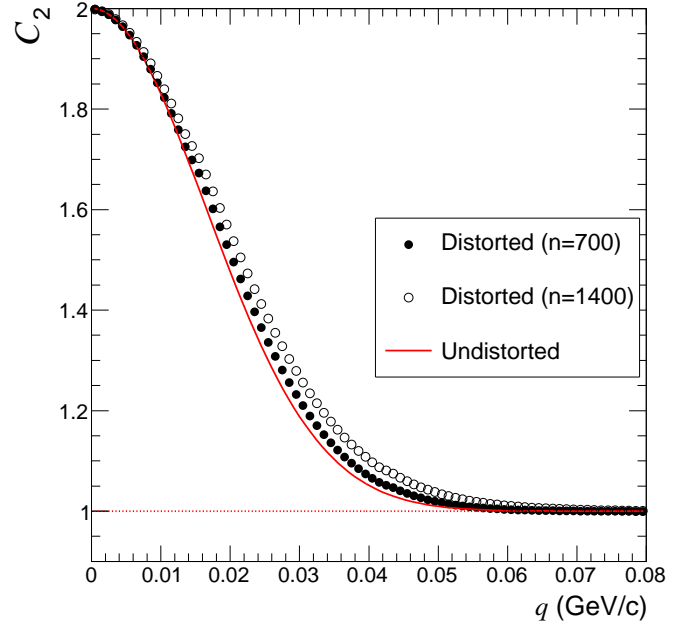}
  \caption{(Color online) Multi-boson distorted 2-pion correlation function compared to the undistorted case.  The setting with $n=700$ roughly correspond to $0-5\%$ Pb--Pb collisions at the LHC.}
  \label{fig:C2multiboson}
\end{figure}

\begin{figure}
\center
  \includegraphics[width=0.49\textwidth]{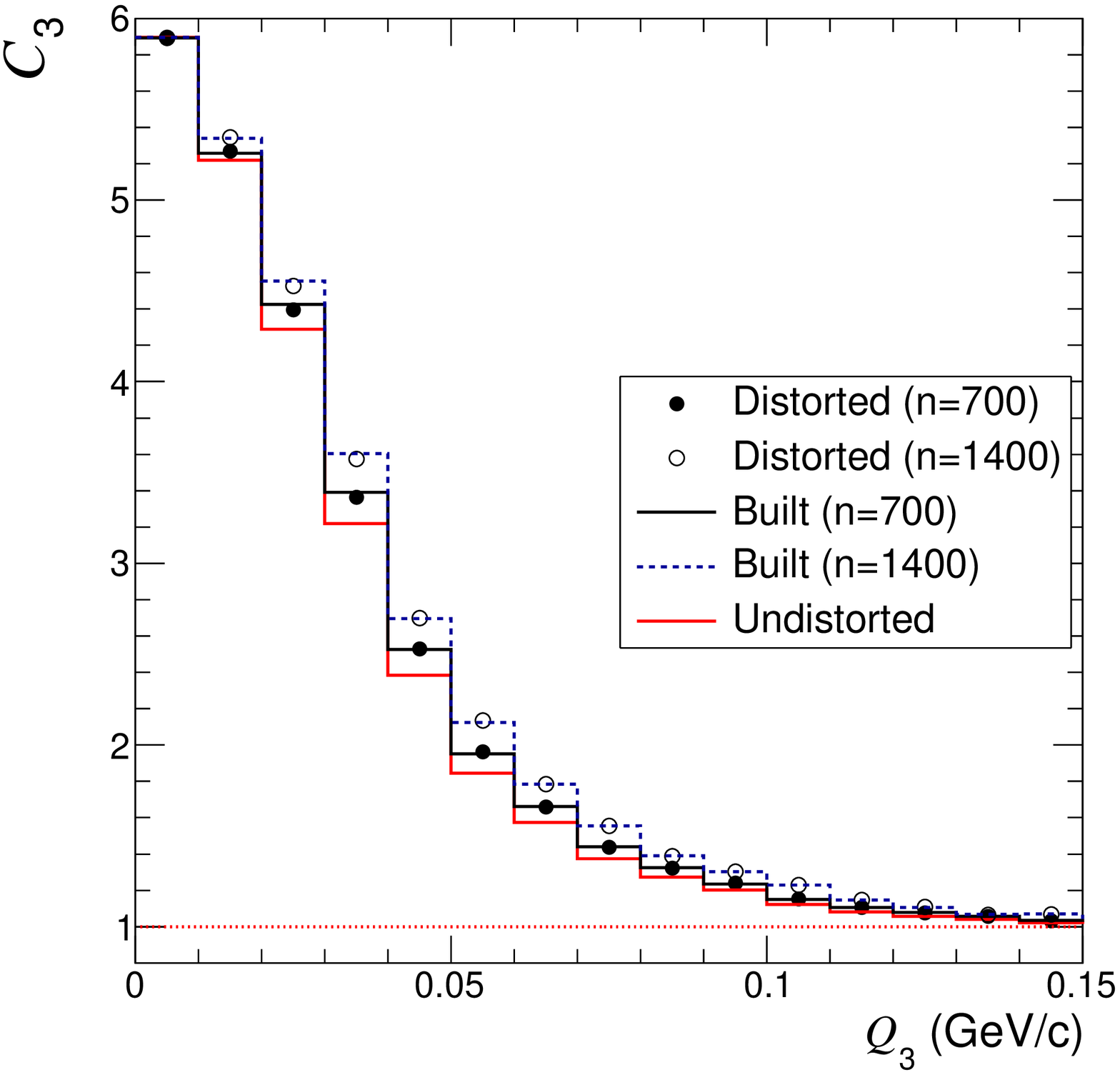}
  \caption{(Color online) The multi-boson distorted 3-pion correlation function compared to the undistorted case.  Also shown is the built correlation function constructed from the distorted 2-pion correlation function in Fig.~\ref{fig:C2multiboson}.  The setting with $n=700$ roughly correspond to $0-5\%$ Pb--Pb collisions at the LHC.}
  \label{fig:C3multiboson}
\end{figure}

\begin{figure}
\center
  \includegraphics[width=0.49\textwidth]{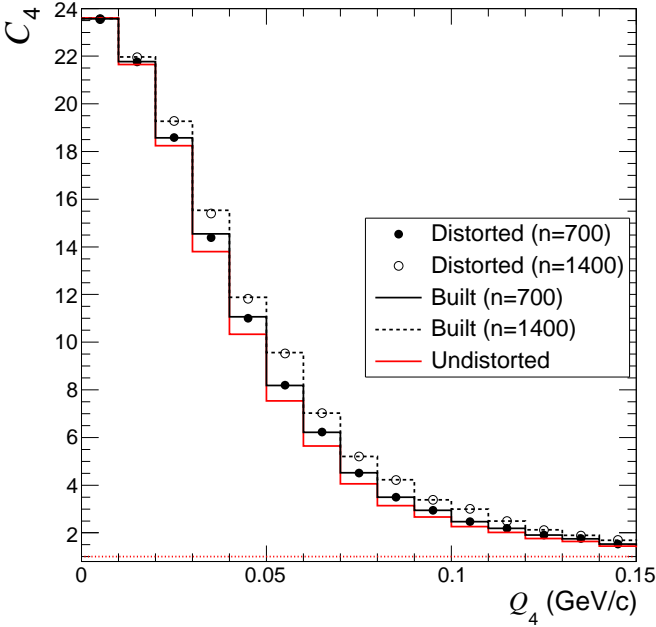}
  \caption{(Color online) The multi-boson distorted 4-pion correlation function compared to the undistorted case.  Also shown is the built correlation function constructed from the distorted 2-pion correlation function in Fig.~\ref{fig:C2multiboson}.  The setting with $n=700$ roughly correspond to $0-5\%$ Pb--Pb collisions at the LHC.}
  \label{fig:C4multiboson}
\end{figure}

In Fig.~\ref{fig:C2multiboson}-\ref{fig:C4multiboson} the distorted and undistorted 2-, 3-, and 4-pion correlation functions are compared. 
It is clear that the multi-boson distortions widen all correlation functions while there is no significant suppression of the intercept, which occurs for much higher pion densities.  
An important feature is the similarity of built and distorted correlation functions.  
This is due to the fact that the 2-pion correlation, which is used in the building procedure, is distorted as well.
The ratio of distorted to built 3- and 4-pion correlation functions stays within 0.01 below unity.
The calculations were also performed with $r_0=7$ fm and $n=700$ which result in larger pion densities.
The increased widening of the 2-pion correlation function leads to a $10\%$ underestimation of $r_0$, when extracted with a Gaussian fit function.
The ratio of distorted to built 3- and 4-pion correlation functions remained very similar.
The proximity of the ratio with unity demonstrates the robustness of the built approach to multi-pion distortions at moderate pion densities.

\section{Summary}
Techniques to isolate and analyze multi-pion QS correlations have been presented. 
The concept of constructed or built correlation functions may be used to estimate the coherent fraction at measured values of relative momentum instead of at the unmeasured intercept of correlation functions.

The distortions from multi-pion symmetrizations at large phase-space densities widens all orders of correlation functions calculated.
The distorted built correlation function is also widened to a very similar extent.
The similarity of the distortions makes the comparison of built and measured correlation functions robust to the multi-pion effects at moderate densities.

The early searches for the DCC have not provided any clear indication for its existence \cite{Aggarwal:1997hd,Brooks:1999xy}.  
In those studies, the main experimental observable considered was event-by-event fluctuations of neutral to charged pion production.  
However, the experimental reconstruction efficiency for $\pi^0$ at low \pt~as compared to charged pions is often far too low to allow for precision measurements of such fluctuations.
As the DCC is expected to radiate at low $\pt$, one also expects an excess in the single-pion spectra at low $\pt$.
Measurements of low $\pt$ pion production in heavy-ion collisions have not revealed a dramatic enhancement \cite{Back:2006tt,Abelev:2012wca} but do not rule out enhancements at the $20\%$ level.

A more promising channel of search for the DCC is one which exploits the expected feature of coherent pion radiation.
Pion coherence suppresses QS correlations and increases substantially for higher order correlation functions, making multi-pion QS correlations a sensitive measure of the DCC.
An additional possible signature of the DCC may lie in the event-by-event $2^{nd}$ order flow vector magnitude, $q_2$, as defined in Refs.~\cite{Voloshin:1994mz,Schukraft:2012ah}.
As the DCC is expected to radiate near the final state where the initial spatial anisotropy in non-central heavy-ion collisions has greatly diminished, one may expect a different azimuthal distribution of coherent pions as compared to chaotic pions.
In particular, one may expect $v_2^{coh} \approx 0$. 
In such a case one expects an anti-correlation between the charged pion $q_2$ and $G$ (coherent fraction).  
For events where the DCC radiates into charged pions, $q_2$ will be diluted.
Whereas for events of neutral pion radiation, $q_2$ will be unchanged.

\newenvironment{acknowledgement}{\relax}{\relax}
\begin{acknowledgement}
\section*{Acknowledgments}
I would like to thank Sergiy Akkelin, Richard Lednick{\' y} and Constantin Loizides for numerous helpful discussions.
This material is based upon work supported in part by the U.S. Department of Energy, Office of Science, Office of Nuclear Physics, under contract number DE-AC02-05CH11231.
\end{acknowledgement}

\appendix

\section{4-pion plane-wave function}
The fully symmetrized 4-pion plane-wave function is given by
{\footnotesize
\begin{eqnarray}
\Psi_4 &=& \sqrt{\frac{1}{24}}[ e^{i(q_{12}'x_{12} + q_{13}'x_{13} + q_{23}'x_{23} + q_{41}'x_{41} + q_{42}'x_{42} + q_{43}'x_{43})/4} \nonumber \\ 
&+& e^{i(q_{12}'x_{21} + q_{13}'x_{23} + q_{23}'x_{13} + q_{41}'x_{42} + q_{42}'x_{41} + q_{43}'x_{43})/4} \nonumber \\ 
&+& {\rm 5\;permutations} \nonumber \\
&+& e^{i(q_{12}'x_{21} + q_{13}'x_{24} + q_{23}'x_{14} + q_{41}'x_{32} + q_{42}'x_{31} + q_{43}'x_{34})/4} \nonumber \\ 
&+& {\rm 2\;permutations} \nonumber \\
&+& e^{i(q_{12}'x_{23} + q_{13}'x_{21} + q_{23}'x_{31} + q_{41}'x_{42} + q_{42}'x_{43} + q_{43}'x_{41})/4} \nonumber \\ 
&+& {\rm 7\;permutations} \nonumber \\
&+& e^{i(q_{12}'x_{41} + q_{13}'x_{42} + q_{23}'x_{12} + q_{41}'x_{34} + q_{42}'x_{31} + q_{43}'x_{32})/4} \nonumber \\ 
&+& {\rm 5\;permutations}],
\end{eqnarray}}
where the second (2-pion), third (2-pair), fourth (3-pion), and fifth (4-pion) plane-wave represent the different symmetrization sequences.

\bibliographystyle{unsrtnat}
\bibliography{biblio}{}

\begin{thebibliography}{46}
\providecommand{\natexlab}[1]{#1}
\providecommand{\url}[1]{\texttt{#1}}
\expandafter\ifx\csname urlstyle\endcsname\relax
  \providecommand{\doi}[1]{doi: #1}\else
  \providecommand{\doi}{doi: \begingroup \urlstyle{rm}\Url}\fi

\bibitem[Goldhaber et~al.(1960)Goldhaber, Goldhaber, Lee, and
  Pais]{Goldhaber:1960sf}
Gerson Goldhaber, Sulamith Goldhaber, Won-Yong Lee, and Abraham Pais.
\newblock {Influence of Bose-Einstein statistics on the anti-proton proton
  annihilation process}.
\newblock \emph{Phys.Rev.}, 120:\penalty0 300--312, 1960.
\newblock \doi{10.1103/PhysRev.120.300}.

\bibitem[Kopylov and Podgoretsky(1975)]{Kopylov:1975rp}
G.I. Kopylov and M.I. Podgoretsky.
\newblock {The interference of two-particle states in particle physics and
  astronomy}.
\newblock \emph{Zh.Eksp.Teor.Fiz.}, 69:\penalty0 414--421, 1975.

\bibitem[Gyulassy et~al.(1979)Gyulassy, Kauffmann, and Wilson]{Gyulassy:1979yi}
M.~Gyulassy, S.K. Kauffmann, and L.W. Wilson.
\newblock {Pion Interferometry of Nuclear Collisions. 1. Theory}.
\newblock \emph{Phys.Rev.}, C20:\penalty0 2267--2292, 1979.
\newblock \doi{10.1103/PhysRevC.20.2267}.

\bibitem[Andreev et~al.(1993)Andreev, Plumer, and Weiner]{Andreev:1992pu}
I.V. Andreev, M.~Plumer, and R.M. Weiner.
\newblock {Quantum statistical approach to Bose-Einstein correlations and its
  experimental implications}.
\newblock \emph{Int.J.Mod.Phys.}, A8:\penalty0 4577--4626, 1993.
\newblock \doi{10.1142/S0217751X93001843}.

\bibitem[Plumer et~al.(1992)Plumer, Razumov, and Weiner]{Plumer:1992au}
M.~Plumer, L.V. Razumov, and R.M. Weiner.
\newblock {Evidence for quantum statistical coherence from experimental data on
  higher order Bose-Einstein correlations}.
\newblock \emph{Phys.Lett.}, B286:\penalty0 335--340, 1992.
\newblock \doi{10.1016/0370-2693(92)91784-7}.

\bibitem[Ornik et~al.(1993)Ornik, Plumer, and Strottmann]{Ornik:1993gb}
U.~Ornik, M.~Plumer, and D.~Strottmann.
\newblock {Bose condensation through resonance decay}.
\newblock \emph{Phys.Lett.}, B314:\penalty0 401--407, 1993.
\newblock \doi{10.1016/0370-2693(93)91257-N}.

\bibitem[Bjorken et~al.(1993)Bjorken, Kowalski, and Taylor]{Bjorken:1993cz}
J.D. Bjorken, K.L. Kowalski, and C.C. Taylor.
\newblock {Baked Alaska}.
\newblock 1993.

\bibitem[Bjorken(1997)]{Bjorken:1997re}
J.D. Bjorken.
\newblock {Disoriented chiral condensate: Theory and phenomenology}.
\newblock \emph{Acta Phys.Polon.}, B28:\penalty0 2773--2791, 1997.

\bibitem[Akkelin et~al.(2002)Akkelin, Lednicky, and Sinyukov]{Akkelin:2001nd}
S.V. Akkelin, R.~Lednicky, and Yu.M. Sinyukov.
\newblock {Correlation search for coherent pion emission in heavy ion
  collisions}.
\newblock \emph{Phys.Rev.}, C65:\penalty0 064904, 2002.
\newblock \doi{10.1103/PhysRevC.65.064904}.

\bibitem[Rajagopal(1997)]{Rajagopal:1997au}
Krishna Rajagopal.
\newblock {Disorienting the chiral condensate at the QCD phase transition}.
\newblock 1997.

\bibitem[Sakurai and Napolitano(2011)]{Sakurai:2011zz}
Jun~John Sakurai and Jim Napolitano.
\newblock {Modern quantum mechanics}.
\newblock 2011.

\bibitem[Zajc(1987)]{Zajc:1986sq}
William~A. Zajc.
\newblock {Monte Carlo Calculational Methods for the Generation of Events with
  Bose-Einstein Correlations}.
\newblock \emph{Phys.Rev.}, D35:\penalty0 3396, 1987.
\newblock \doi{10.1103/PhysRevD.35.3396}.

\bibitem[Pratt(1993)]{Pratt:1993uy}
S.~Pratt.
\newblock {Pion lasers from high-energy collisions}.
\newblock \emph{Phys.Lett.}, B301:\penalty0 159--164, 1993.
\newblock \doi{10.1016/0370-2693(93)90682-8}.

\bibitem[Pratt(1994)]{Pratt:1994cg}
S.~Pratt.
\newblock {Deciphering the CENTAURO puzzle}.
\newblock \emph{Phys.Rev.}, C50:\penalty0 469--479, 1994.
\newblock \doi{10.1103/PhysRevC.50.469}.

\bibitem[Lednicky et~al.(2000)Lednicky, Lyuboshitz, Mikhailov, Sinyukov,
  Stavinsky, et~al.]{Lednicky:1999xz}
R.~Lednicky, V.~Lyuboshitz, K.~Mikhailov, Yu. Sinyukov, A.~Stavinsky, et~al.
\newblock {Multiboson effects in multiparticle production}.
\newblock \emph{Phys.Rev.}, C61:\penalty0 034901, 2000.
\newblock \doi{10.1103/PhysRevC.61.034901}.

\bibitem[Zhang et~al.(1998)Zhang, Scotto, and Heinz]{Zhang:1998sz}
Q.H. Zhang, P.~Scotto, and Ulrich~W. Heinz.
\newblock {Multiboson effects and the normalization of the two pion correlation
  function}.
\newblock \emph{Phys.Rev.}, C58:\penalty0 3757--3760, 1998.
\newblock \doi{10.1103/PhysRevC.58.3757}.

\bibitem[Csorgo et~al.(1996)Csorgo, Lorstad, and Zimanyi]{Csorgo:1994in}
T.~Csorgo, B.~Lorstad, and J.~Zimanyi.
\newblock {Bose-Einstein correlations for systems with large halo}.
\newblock \emph{Z.Phys.}, C71:\penalty0 491--497, 1996.
\newblock \doi{10.1007/s002880050195}.

\bibitem[Heinz and Zhang(1997)]{Heinz:1997mr}
Ulrich~W. Heinz and Q.H. Zhang.
\newblock {What can we learn from three pion interferometry?}
\newblock \emph{Phys.Rev.}, C56:\penalty0 426--431, 1997.
\newblock \doi{10.1103/PhysRevC.56.426}.

\bibitem[Csorgo(2002)]{Csorgo:1999sj}
T.~Csorgo.
\newblock {Particle interferometry from 40-MeV to 40-TeV}.
\newblock \emph{Heavy Ion Phys.}, 15:\penalty0 1--80, 2002.
\newblock \doi{10.1556/APH.15.2002.1-2.1}.

\bibitem[Heinz and Sugarbaker(2004)]{Heinz:2004pv}
Ulrich~W. Heinz and Alex Sugarbaker.
\newblock {Projected three-pion correlation functions}.
\newblock \emph{Phys.Rev.}, C70:\penalty0 054908, 2004.
\newblock \doi{10.1103/PhysRevC.70.054908}.

\bibitem[Abelev et~al.(2014{\natexlab{a}})]{Abelev:2013pqa}
Betty Abelev et~al.
\newblock {Two and Three-Pion Quantum Statistics Correlations in Pb--Pb
  Collisions at $\sqrt{s_{NN}}=2.76$ TeV at the LHC}.
\newblock \emph{Phys.Rev.}, C89:\penalty0 024911, 2014{\natexlab{a}}.
\newblock \doi{10.1103/PhysRevC.89.024911}.

\bibitem[Aamodt et~al.(2010)]{Aamodt:2010jj}
K~Aamodt et~al.
\newblock {Two-pion Bose-Einstein correlations in pp collisions at
  $\sqrt{s}=900$ GeV}.
\newblock \emph{Phys.Rev.}, D82:\penalty0 052001, 2010.
\newblock \doi{10.1103/PhysRevD.82.052001}.

\bibitem[Aamodt et~al.(2011{\natexlab{a}})]{Aamodt:2011kd}
K.~Aamodt et~al.
\newblock {Femtoscopy of pp collisions at $\sqrt{s}=0.9$ and 7 TeV at the LHC
  with two-pion Bose-Einstein correlations}.
\newblock \emph{Phys.Rev.}, D84:\penalty0 112004, 2011{\natexlab{a}}.
\newblock \doi{10.1103/PhysRevD.84.112004}.

\bibitem[Abelev et~al.(2014{\natexlab{b}})]{Abelev:2014pja}
Betty~Bezverkhny Abelev et~al.
\newblock {Freeze-out radii extracted from three-pion cumulants in pp, p–Pb
  and Pb–Pb collisions at the LHC}.
\newblock \emph{Phys.Lett.}, B739:\penalty0 139--151, 2014{\natexlab{b}}.
\newblock \doi{10.1016/j.physletb.2014.10.034}.

\bibitem[Csorgo and Hegyi(2000)]{Csorgo:2000pf}
T.~Csorgo and S.~Hegyi.
\newblock {Model independent shape analysis of correlations in 1, 2 or 3
  dimensions}.
\newblock \emph{Phys.Lett.}, B489:\penalty0 15--23, 2000.
\newblock \doi{10.1016/S0370-2693(00)00935-7}.

\bibitem[Ackerstaff et~al.(1998)]{Ackerstaff:1998py}
K.~Ackerstaff et~al.
\newblock {Bose-Einstein correlations of three charged pions in hadronic Z0
  decays}.
\newblock \emph{Eur.Phys.J.}, C5:\penalty0 239--248, 1998.
\newblock \doi{10.1007/s100520050265}.

\bibitem[Abreu et~al.(1995)]{Abreu:1995sq}
P.~Abreu et~al.
\newblock {Observation of short range three particle correlations in e+ e-
  annihilations at LEP energies}.
\newblock \emph{Phys.Lett.}, B355:\penalty0 415--424, 1995.
\newblock \doi{10.1016/0370-2693(95)00711-S}.

\bibitem[Achard et~al.(2002)]{Achard:2002ja}
P.~Achard et~al.
\newblock {Measurement of genuine three particle Bose-Einstein correlations in
  hadronic $Z$ decay}.
\newblock \emph{Phys.Lett.}, B540:\penalty0 185--198, 2002.
\newblock \doi{10.1016/S0370-2693(02)02142-1}.

\bibitem[Sinyukov et~al.(1998)Sinyukov, Lednicky, Akkelin, Pluta, and
  Erazmus]{Sinyukov:1998fc}
Yu. Sinyukov, R.~Lednicky, S.V. Akkelin, J.~Pluta, and B.~Erazmus.
\newblock {Coulomb corrections for interferometry analysis of expanding hadron
  systems}.
\newblock \emph{Phys.Lett.}, B432:\penalty0 248--257, 1998.
\newblock \doi{10.1016/S0370-2693(98)00653-4}.

\bibitem[Lednicky(2009)]{Lednicky:2005tb}
Richard Lednicky.
\newblock {Finite-size effects on two-particle production in continuous and
  discrete spectrum}.
\newblock \emph{Phys.Part.Nucl.}, 40:\penalty0 307--352, 2009.
\newblock \doi{10.1134/S1063779609030034}.

\bibitem[Alt and Mukhamedzhanov(1993)]{Alt:1992js}
E.O. Alt and A.M. Mukhamedzhanov.
\newblock {On the asymptotic solution of the Schrodinger equation for three
  charged particles}.
\newblock \emph{Phys.Rev.}, A47:\penalty0 2004--2022, 1993.
\newblock \doi{10.1103/PhysRevA.47.2004}.

\bibitem[Lednicky and Amelin(1995)]{LednickyAmelin}
R.~Lednicky and N.~Amelin.
\newblock {Bose-Einstein correlations and classical transport models}.
\newblock page~34, 1995.
\newblock \doi{SUBATECH-95-08}.

\bibitem[Alt et~al.(1999)Alt, Csorgo, Lorstad, and
  Schmidt-Sorensen]{Alt:1998nr}
E.O. Alt, T.~Csorgo, B.~Lorstad, and J.~Schmidt-Sorensen.
\newblock {Coulomb corrections to the three-body correlation function in
  high-energy heavy-ion reactions}.
\newblock \emph{Phys.Lett.}, B458:\penalty0 407--414, 1999.
\newblock \doi{10.1016/S0370-2693(99)00588-2}.

\bibitem[Alt et~al.(2005)Alt, Irgaziev, and Mukhamedzhanov]{Alt:2005su}
E.O. Alt, B.F. Irgaziev, and A.M. Mukhamedzhanov.
\newblock {Three-body Coulomb final-state interaction effects in the Coulomb
  breakup of light nuclei}.
\newblock \emph{Mod.Phys.Lett.}, A20:\penalty0 947--964, 2005.
\newblock \doi{10.1142/S0217732305017378}.

\bibitem[Lednicky and Podgoretsky(1979)]{Lednicky:1979ig}
R.~Lednicky and M.I. Podgoretsky.
\newblock {The interference of identical particles emitted by sources of
  different sizes}.
\newblock \emph{Sov.J.Nucl.Phys.}, 30:\penalty0 432, 1979.

\bibitem[Csorgo et~al.(2004)Csorgo, Hegyi, and Zajc]{Csorgo:2003uv}
T.~Csorgo, S.~Hegyi, and W.A. Zajc.
\newblock {Bose-Einstein correlations for Levy stable source distributions}.
\newblock \emph{Eur.Phys.J.}, C36:\penalty0 67--78, 2004.
\newblock \doi{10.1140/epjc/s2004-01870-9}.

\bibitem[Adams et~al.(2005)]{Adams:2004yc}
J.~Adams et~al.
\newblock {Pion interferometry in Au+Au collisions at $\snn=200$ GeV}.
\newblock \emph{Phys.Rev.}, C71:\penalty0 044906, 2005.
\newblock \doi{10.1103/PhysRevC.71.044906}.

\bibitem[Aamodt et~al.(2011{\natexlab{b}})]{Aamodt:2011mr}
K.~Aamodt et~al.
\newblock {Two-pion Bose-Einstein correlations in central PbPb collisions at
  $\snn=2.76$ TeV}.
\newblock \emph{Phys.Lett.}, B696:\penalty0 328--337, 2011{\natexlab{b}}.
\newblock \doi{10.1016/j.physletb.2010.12.053}.

\bibitem[Kisiel et~al.(2006)Kisiel, Taluc, Broniowski, and
  Florkowski]{Kisiel:2005hn}
Adam Kisiel, Tomasz Taluc, Wojciech Broniowski, and Wojciech Florkowski.
\newblock {THERMINATOR: THERMal heavy-IoN generATOR}.
\newblock \emph{Comput.Phys.Commun.}, 174:\penalty0 669--687, 2006.
\newblock \doi{10.1016/j.cpc.2005.11.010}.

\bibitem[Chojnacki et~al.(2012)Chojnacki, Kisiel, Florkowski, and
  Broniowski]{Chojnacki:2011hb}
Mikolaj Chojnacki, Adam Kisiel, Wojciech Florkowski, and Wojciech Broniowski.
\newblock {THERMINATOR 2: THERMal heavy IoN generATOR 2}.
\newblock \emph{Comput.Phys.Commun.}, 183:\penalty0 746--773, 2012.
\newblock \doi{10.1016/j.cpc.2011.11.018}.

\bibitem[Aggarwal et~al.(1998)]{Aggarwal:1997hd}
M.M. Aggarwal et~al.
\newblock {Search for disoriented chiral condensates in 158-GeV/A Pb + Pb
  collisions}.
\newblock \emph{Phys.Lett.}, B420:\penalty0 169--179, 1998.
\newblock \doi{10.1016/S0370-2693(97)01528-1}.

\bibitem[Brooks et~al.(2000)]{Brooks:1999xy}
T.C. Brooks et~al.
\newblock {A Search for disoriented chiral condensate at the Fermilab
  Tevatron}.
\newblock \emph{Phys.Rev.}, D61:\penalty0 032003, 2000.
\newblock \doi{10.1103/PhysRevD.61.032003}.

\bibitem[Back et~al.(2007)]{Back:2006tt}
B.B. Back et~al.
\newblock {Identified hadron transverse momentum spectra in Au+Au collisions at
  $\sqrt{s_{NN}}=62.4$ GeV}.
\newblock \emph{Phys.Rev.}, C75:\penalty0 024910, 2007.
\newblock \doi{10.1103/PhysRevC.75.024910}.

\bibitem[Abelev et~al.(2012)]{Abelev:2012wca}
Betty Abelev et~al.
\newblock {Pion, Kaon, and Proton Production in Central Pb--Pb Collisions at
  $\sqrt{s_{NN}} = 2.76$ TeV}.
\newblock \emph{Phys.Rev.Lett.}, 109:\penalty0 252301, 2012.
\newblock \doi{10.1103/PhysRevLett.109.252301}.

\bibitem[Voloshin and Zhang(1996)]{Voloshin:1994mz}
S.~Voloshin and Y.~Zhang.
\newblock {Flow study in relativistic nuclear collisions by Fourier expansion
  of Azimuthal particle distributions}.
\newblock \emph{Z.Phys.}, C70:\penalty0 665--672, 1996.
\newblock \doi{10.1007/s002880050141}.

\bibitem[Schukraft et~al.(2013)Schukraft, Timmins, and
  Voloshin]{Schukraft:2012ah}
Jurgen Schukraft, Anthony Timmins, and Sergei~A. Voloshin.
\newblock {Ultra-relativistic nuclear collisions: event shape engineering}.
\newblock \emph{Phys.Lett.}, B719:\penalty0 394--398, 2013.
\newblock \doi{10.1016/j.physletb.2013.01.045}.

\end{thebibliography}

\end{document}